\documentclass[12pt,subeqn,a4paper]{article}

\usepackage{amssymb,amsmath,amsfonts,amsthm, amscd, mathrsfs,helvet}
\usepackage{bm}
\usepackage{graphicx,verbatim}
\usepackage{psfrag}
\usepackage[all]{xy}

\def\r2{\sqrt{2}}

\def\bea{\begin{eqnarray} }
\def\eea{\end{eqnarray}}
\def\be{\begin{equation} }
\def\ee{\end{equation}}
\def\nn{\nonumber}

\begin{document}
\newcommand{\nd}[1]{/\hspace{-0.5em} #1}
\begin{titlepage}
\begin{flushright}
{\bf December 2008} \\ 
%DAMTP-\\
%SWAT-\\ 
%hep-th/yymmnnn \\
\end{flushright}
\begin{centering}
\vspace{.2in}
 {\large {\bf Spiky Strings and Spin Chains}}

\vspace{.3in}

Nick Dorey and Manuel Losi\\
\vspace{.1 in}
DAMTP, Centre for Mathematical Sciences \\ 
University of Cambridge, Wilberforce Road \\ 
Cambridge CB3 0WA, UK \\
\vspace{.2in}
%and \\ 
%\vspace{.2in}
%
%
\vspace{.4in}
{\bf Abstract} \\
\end{centering}
We determine spectral curves for the known spiky string solutions  
in $AdS$ space in the limit of large angular momentum. 
We also construct generic multi-spike 
solutions in this limit and compute the corresponding 
spectral data. The resulting spectral curves precisely match those of
the classical spin chain describing the dual operators in one-loop
gauge theory. Our results confirm the map between string
theory and gauge theory degrees of freedom proposed in 
{\tt arXiv:0805.4387 [hep-th]}.  
 
%\vspace{.05in}
%\baselineskip=.3in
\end{titlepage}
\paragraph{}
\section{Introduction}
\paragraph{}
The AdS/CFT correspondence predicts an exact equivalence between the
operator dimensions of planar ${\cal N}=4$ SUSY Yang-Mills and the
energy levels of free string theory on $AdS_{5}\times S^{5}$. This
matching is complicated by the fact that the two theories involved are
only tractable at small and large values of the 't Hooft coupling, 
$\lambda=g^{2}N$ respectively. Nevertheless, as we vary $\lambda$ 
from small to large values, each local operator of definite
scaling dimension in perturbative gauge theory should go over to a
particular string state on $AdS_{5}\times S^{5}$. The emergence of
integrability on both sides of the correspondence \cite{MZ,B,BPR} 
leads to the hope
that we can describe this interpolation exactly for generic states.  
So far this has only been accomplished definitively for 
operators with very large $R$-charge where a description of 
the spectrum in terms of asymptotic states can be obtained on both
sides. Analytic results for the dispersion relation \cite{BDS,Beis} 
and scattering matrix \cite{Beis,MS,BS1,BES} 
of these states interpolate smoothly between weakly coupled
gauge theory and semiclassical string theory, providing a complete
description of the spectrum in this limit. In a recent paper
\cite{D1}, one of the present authors proposed a 
similarly precise matching between states in
the limit of large conformal spin (denoted $S$). The character of 
this limit is quite different from the case of large $R$-charge where 
gauge theory operators are described by a quantum spin chain of 
infinite length. Instead, large $S$ leads to a {\em classical} spin chain of 
{\em fixed length} on the gauge theory side. 
In \cite{D1}, the spectrum of semiclassical string
theory in this limit was shown to exactly match that of the
corresponding sector of one-loop gauge theory up to a single overall
normalisation. A precise map between string theory and gauge theory
degrees of freedom was also proposed. In this paper we will expand on
the proposal of \cite{D1} and perform several detailed tests. In
particular we will construct explicit string solutions in the
large-$S$ limit and check the correspondence to one-loop gauge theory
directly. The main results are described
in the remainder of this introductory section.
\paragraph{}
As in \cite{D1}, we focus on single-trace operators of the form,   
\bea 
\hat{O} & \sim & {\rm Tr}_{N}\left[\,\mathcal{D}_{+}^{s_{1}}Z\,
\mathcal{D}_{+}^{s_{2}}Z\,\ldots\, \mathcal{D}_{+}^{s_{J}}Z\,\right] 
\label{sl2} 
\eea
having total Lorentz spin  $S = \sum_{l=1}^{J}\,\,s_{l}$ and twist
(or equivalently $R$-charge) equal to $J$.  
Here $\mathcal{D}_{+}$ is a covariant derivative with conformal spin
plus one and $Z$ is one of the three complex adjoint scalar fields of the
$\mathcal{N}=4$ theory. These operators belong to the non-compact $sl(2)$
sector of the ${\cal N}=4$ theory. In this sector, the one-loop dilatation
operator is equivalent to the Hamiltonian of an integrable
$SL(2,\mathbb{R})$ spin chain of length $J$ \cite{MZ,B,KK}
which can be diagonalised exactly using the Bethe ansatz. 
Here, we are primarily interested in
the large spin limit\footnote{For other relevant work on this limit
  see \cite{Kruc,Kruc2,Stef,MTT,Sakai}} $S\rightarrow \infty$ with fixed $J$, 
where the non-compact spins
at each site are highly excited and can be replaced by corresponding 
classical variables \cite{BGK1,BBGK,BGK2}. 
The phase space of the resulting classical spin
chain splits into sectors labelled by a positive integer $K\leq J$. In
each sector the phase space is characterised by a spectral curve,          
\bea 
\Gamma_{K} \,:\qquad{} t\,\,+\,\,\frac{1}{t} & = & \,\, 2\,\,+\,\,\frac{q_{2}}
{{u}^{2}}\,\,+\,\, \frac{{q}_{3}}
{{u}^{3}}\,\,+\,\,\ldots\,\,+\,\, \frac{{q}_{K}}
{{u}^{K}} \qquad{}  \label{crv1} \eea 
which defines a Riemann surface of genus $K-2$. 
The moduli, ${q}_{j}$ $j=2,\ldots, K$, of the curve 
correspond to the higher conserved charges of the spin chain. 
The one-loop anomalous dimensions of operators in this sector are
given as, 
\bea  
\Delta-S-J & = & \frac{\lambda}{4\pi^{2}}
\left( \log {q}_{K}\,\,+\,\,C_{\rm 1-loop}\,\,+\,\,
O\left(\frac{1}{\log^{2}S}\right) \right) \label{gpred} \eea 
where $C_{\rm 1-loop}$ is an undetermined constant which may depend on
$K$ and $J$ but not on the charges $q_{k}$. 
As $q_{K}\sim S^{K}$, this formula exhibits the characteristic $K\log S$
scaling of the anomalous dimensions of twist $K$ operators \cite{GW, K1}.  
As reviewed in Section 2 below, 
semi-classical quantisation leads to a discrete 
spectrum for the charges $q_{k}$ and hence also for the anomalous
dimensions.   
\paragraph{}  
For large 't Hooft coupling, operators of the form (\ref{sl2}) are
dual to semiclassical strings moving on an $AdS_{3}\times S^{1}$ 
submanifold of $AdS_{5}\times S^{5}$. Here the spin $S$ corresponds to angular
momentum on $AdS_{3}$ and twist $J$ to angular momentum on $S^{1}$. 
The large-$S$ limit of the string spectrum was studied in \cite{D1} and found
to be identical to the semiclassical spectrum of the spin chain
described above up to a single coupling-dependent normalisation. The
equality of the two spectra was demonstrated using the finite-gap
formalism for classical string theory developed in \cite{KMMZ,KZ} in
which each classical solution of string theory in $AdS_{3}\times S^{1}$ has
an associated spectral curve which encodes the values of the higher 
conserved charges of the worldsheet sigma model.
It was shown that the spectral curve of the ``$K$-gap''
string solutions reduces to the gauge theory curve $\Gamma_{K}$ in the
large-$S$ limit. The symplectic form and Hamiltonian were also found
to coincide with their gauge theory counterparts in this limit.
In particular, the leading semiclassical 
spectrum of string energies is given by, 
\bea  
\Delta-S-J & = & \frac{\sqrt{\lambda}}{2\pi}
\left( \log {q}_{K}\,\,+\,\,C_{\rm string}\,\,+\,\,
O\left(\frac{1}{\log^{2}S}\right) \right) \label{strpred} \eea
where $C_{\rm string}$ is a constant which we will determine below. 
A more specific proposal for how gauge theory and string theory
degrees of freedom are related was also made in \cite{D1}. The key
idea was that generic $K$-gap string solutions should develop $K$ cusps
which approach the boundary as $S\rightarrow \infty$. 
Here we will make these ideas more concrete by constructing the
limiting string solutions explicitly and checking their correspondence
to the gauge theory spin chain.  
\paragraph{}
We begin by studying the known large-spin solutions of string theory
on $AdS_{3}$ (motion on $S^{1}$ can be neglected in this limit). These
are the multiply folded spinning string of Gubser, Klebanov and Polyakov
(GKP) and the spiky string of Kruczenski. Both these solutions have 
cusps which approach the boundary as $S\rightarrow \infty$. In
this limit, the charge density corresponding to the spin $S$ becomes
$\delta$-function localised at the spikes. We
calculate the monodromy matrix of each solution and thereby determine
the corresponding spectral curve. The resulting curves for the $K$
spike solutions correspond to
particular points in the moduli space of the gauge theory curve
$\Gamma_{K}$. For both types of solution, these turn out to be special
points in the moduli space where $\Gamma_{K}$ degenerates from genus
$K-2$ to genus zero. The filling numbers of Bethe roots for the dual
gauge theory operator are evaluated from the curve in each case. The
two types of solution correspond to particular single-cut
configurations in gauge theory.  
\paragraph{}
In the next part of the paper we construct a $(K-1)$-parameter
generalisation of the spiky string solution of Kruczenski \cite{Kruc}.
In the solution of \cite{Kruc}, the angular separation between each
pair of adjacent cusps is the same. Here, we construct a limiting
solution at large-$S$ in 
which the angular separations $\Delta\theta_{j}=\theta_{j}-\theta_{j-1}$, with 
$j=1,2,\ldots, K$, between adjacent
cusps can be adjusted subject 
to the constraints, 
\bea 
0<\Delta\theta_{j}\leq \pi & \qquad{} & \sum_{j=1}^{K}\,
\Delta\theta_{j}\,=\, 2\pi n \nn \eea
for positive integer $n$ (see Figure 1). 
\begin{figure}
\centering
\psfrag{a}{\footnotesize{$\Delta\theta_1$}}
\psfrag{b}{\footnotesize{$\Delta\theta_2$}}
\psfrag{c}{\footnotesize{$\Delta\theta_3$}}
%\psfrag{d}{\footnotesize{$\ldots$}}
%\psfrag{e}{\footnotesize{$ $}}
\psfrag{f}{\footnotesize{$\Delta\theta_n$}}
\includegraphics[width=100mm]{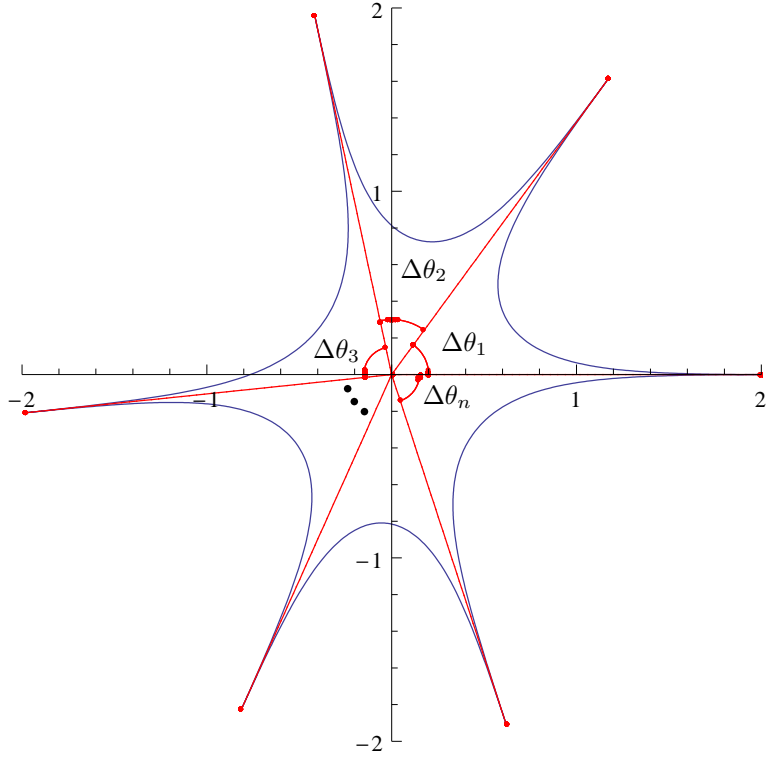}
\caption{The general multi-spike solution.}
\label{Intro_spike}
\end{figure}
The original solution \cite{Kruc} corresponds to the choice
$\theta_{j}=2\pi/K$ for all $j=1,2,\ldots, K$ where the spikes lie at
the vertices of a regular $K$-gon. The $N$-folded string of GKP also
arises as a special case corresponding to the choice $\Delta
\theta_{j}=\pi$ for all $j=1,2,\ldots, K$ with $K=2N$. 
\paragraph{}
The next step is to 
evaluate the monodromy matrix and spectral curve of the limiting
solution described above. Once again the curve precisely matches the
gauge theory result (\ref{crv1}). The conserved charges ${q}_{k}$
are related to the initial angular positions $\theta_{i}$ of the cusps as, 
\bea 
q_{k} & = & \left(\frac{-2S}{K}\right)^{k} 
\sum_{1\leq j_{1}<\ldots <j_{k}\leq K}\, \prod_{l=1}^{k} \,\,\sin
\left(\frac{\theta_{j_{l+1}}-\theta_{j_{l}}}{2}\right) \nn \eea
for $k=1,2,\ldots,K$. We also evaluate the string energy
which obeys the expected relation (\ref{strpred}) with constant part
given as, 
\bea
C_{\rm string} & = &
K\left[\log\left(\frac{8\pi}{\sqrt{\lambda}}\right) \,\,-\,\,1 \right]
+\log(-1)^{K+n} \nn \eea 
\paragraph{}
All the solutions studied in this paper exhibit the localisation 
of angular momentum density predicted in \cite{D1}. In each case the
worldsheet charge density $j_{\tau}$ at large $S$ has the form,  
\bea 
j_{\tau}(\sigma,\tau) & \simeq & 
\frac{8\pi}{\sqrt{\lambda}} \sum_{k=0}^{K-1} L_k \delta(\sigma -
\sigma_k) \nn 
\eea
for some $\mathfrak{su}(1,1)$-valued variables $L_{k}$, where
$\sigma=\sigma_{k}$ are the locations of the $K$ cusps of the string. 
In \cite{D1}, it was proposed that these variables should be identified
directly with the classical spin variables of the one-loop spin
chain of length $K$. The agreement of the string theory and gauge
theory curves described above provides further evidence for this
identification. We also calculate the 
values of the variables $L_{k}$ for the general spiky string.  

\section{The gauge theory spin chain}
\paragraph{}
We consider the one-loop anomalous dimensions of operators in the  
non-compact rank one subsector of planar $\mathcal{N}=4$ SUSY Yang-Mills 
(also known as the $sl(2)$ sector). Generic single-trace operators in
this sector are labelled by their Lorentz spin $S$ and $U(1)_{R}$
charge $J$ and have the form (\ref{sl2}).  The classical dimension 
of each operator is $\Delta_{0}=J+S$ and its twist (classical
dimension minus spin) is therefore equal to $J$. 
\paragraph{}
The one-loop anomalous dimensions of operators in the $sl(2)$ sector
are determined by the eigenvalues of the Hamiltonian of the 
Heisenberg XXX$_{-\frac{1}{2}}$ spin chain with $J$ sites. Each site
of this chain 
carries a representation of $SL(2,\mathbb{R})$ with quadratic Casimir
equal to minus one half. Our discussion of the chain in this section
follows that of \cite{BGK1,BGK2} (See in particular Section 2.2 of
\cite{BGK2}). 
\paragraph{}
Here we will focus on the large-spin limit of
the chain: $S\rightarrow \infty$ with $J$ fixed. This is
effectively a semiclassical limit where $1/S$ plays the role of
Planck's constant $\hbar$ \cite{BGK1,BGK2}. In this limit the 
quantum spins are replaced by the classical variables 
$\mathcal{L}^{\pm}_{k}$, $\mathcal{L}^{0}_{k}$, for $k=1,2,\ldots, J$, 
introduced above. The commutators of spin operators are
replaced by the Poisson brackets 
\bea 
\{ \mathcal{L}^{+}_{k}, \mathcal{L}^{-}_{k'}\} \,=\, 
2i\delta_{kk'} \mathcal{L}^{0}_{k} & \qquad{} & 
\{ \mathcal{L}^{0}_{k}, \mathcal{L}^{\pm}_{k'}\}
\,=\, \pm i\delta_{kk'} 
\mathcal{L}^{\pm}_{k} \label{pb} \eea
of these classical spins. 
As the quadratic Casimir
equal to $-1/2$ is negligible in the $S\rightarrow \infty$ limit, 
the classical spins at
each site obey the relation
\bea 
\mathcal{L}^{+}_{k}\,\mathcal{L}^{-}_{k}\,\,+\,\, \left(
\mathcal{L}^{0}_{k}\right)^{2} & = & 0 \label{Cas} \eea 
up to $1/S$ corrections. 
We will restrict our attention to states obeying the highest weight
condition, 
\bea 
\sum_{k=1}^{J} \mathcal{L}^{\pm}_{k} & = &  0  
\label{hw} \eea 
\paragraph{}
Integrability of the classical spin chain starts from the existence 
of a Lax matrix, 
\bea 
\mathbb{L}_{k}(u) & = & \left(\begin{array}{cc} u+
i\mathcal{L}^{0}_{k} & i\mathcal{L}^{+}_{k} \\ i\mathcal{L}^{-}_{k}
& u-i\mathcal{L}^{0}_{k} \end{array}\right) 
\label{lax} \eea
where $u\in \mathbb{C}$ is a spectral parameter.  
A tower of conserved quantities are obtained by constructing the monodromy, 
\bea 
t_{J}(u) & = & {\rm
  tr}_{2}\left[\mathbb{L}_{1}(u)\mathbb{L}_{2}(u)
\ldots\mathbb{L}_{J}(u)\right]   \nn \\ 
& = & 2u^{J}\,+\, q_{2}u^{J-2} \,+
\,\ldots\,+\,q_{J-1}u\, +\, q_{J} \label{monod3} \eea
At large-$S$ we find 
$q_{2}=-S^{2}$ up to corrections of order $1/S$. 
One may check starting from the Poisson brackets (\ref{pb}) that the
conserved charges, $q_{j}$, $j=2,3,\ldots J$ are in involution: 
$\{q_{j}, \,\, q_{k}\}=0$ $\forall\, j,\,k$. Taking into account the
highest-weight constraint (\ref{hw}), this is a sufficient number of
conserved quantities for complete integrability of the chain. 
\paragraph{}
The one-loop spectrum of operator
dimensions at large-$S$ is determined from the semiclassical spectrum of
the spin chain. It has different branches, labelled by an integer
$K\leq J$, corresponding to the highest
non-zero conserved charge \cite{BGK2}, 
\bea 
q_{K}\neq 0 & \qquad{} & q_{j}=0\qquad{}\forall\,\,j>K \nn \eea 
The one-loop anomalous dimensions in this sector are given as, 
\bea  
\Delta-S-J & = & \frac{\lambda}{4\pi^{2}}
\left( \log {q}_{K}\,\,+\,\,C_{\rm 1-loop}\,\,+\,\,
O\left(\frac{1}{\log^{2}S}\right) \right) \qquad{} (\ref{gpred}) \nn \eea
where $C_{\rm 1-loop}$ is an undetermined constant which is
independent of the moduli $q_{j}$. 
We call the branch with $K=J$ the highest sector. For each  
$K<J$ there is also a sector of states 
isomorphic to the highest sector of a shorter chain with only $K$
sites. 
In the limit of large $S$, the conserved charge $q_{j}$ scales as
$S^{j}$ for $j=2,\ldots,K$. 
\paragraph{}
At the classical level, the conserved charges $q_{j}$ vary
continuously. The leading large-$S$ form of the discrete spectrum 
arises from imposing appropriate Bohr-Sommerfeld quantisation
conditions. These conditions are formulated in terms of the {\it
  spectral curve}, 
\bea 
\Gamma_{K} \,:\qquad{} t\,\,+\,\,\frac{1}{t} & = & \,\, 2\,\,+\,\,\frac{q_{2}}
{{u}^{2}}\,\,+\,\, \frac{{q}_{3}}
{{u}^{3}}\,\,+\,\,\ldots\,\,+\,\, \frac{{q}_{K}}
{{u}^{K}} \qquad{} \qquad{} (\ref{crv1}) \nn \eea 
which corresponds to a double cover of the 
$u$ plane with $2K-2$ branch points at $u=u_{1},\ldots,u_{2K-2}$ as
shown in Figure 2. We also define $K-1$ one-cycles $\alpha_{j}$,
$j=1,\ldots, K-1$ as shown in the Figure\footnote{To state the main
  results of \cite{BGK1, BGK2}, we will not need a
to introduce a full basis of cycles on $\Gamma_{K}$.}.
\begin{figure}
\centering
\psfrag{x}{\footnotesize{$u$}}
\psfrag{m1}{\footnotesize{$-1$}}
\psfrag{p1}{\footnotesize{$+1$}}
\psfrag{C1m}{\footnotesize{$C_{1}^{-}$}}
\psfrag{C1p}{\footnotesize{$C_{1}^{+}$}}
\psfrag{C1k}{\footnotesize{$C_{K/2}^{-}$}}
\psfrag{C1j}{\footnotesize{$C_{K/2}^{+}$}}
\psfrag{a}{\footnotesize{$u_{2K-2}$}}
\psfrag{b}{\footnotesize{$u_{2K-3}$}}
\psfrag{c}{\footnotesize{$u_{4}$}}
\psfrag{d}{\footnotesize{$u_{3}$}}
\psfrag{e}{\footnotesize{$u_{2}$}} 
\psfrag{f}{\footnotesize{$u_{1}$}}
\psfrag{A1}{\footnotesize{$\alpha_{1}$}}
\psfrag{A2}{\footnotesize{$\alpha_{2}$}}
\psfrag{A3}{\footnotesize{$\alpha_{K-1}$}}
\psfrag{h}{\footnotesize{$a^{(K-1)}_{+}$}}
\includegraphics[width=100mm]{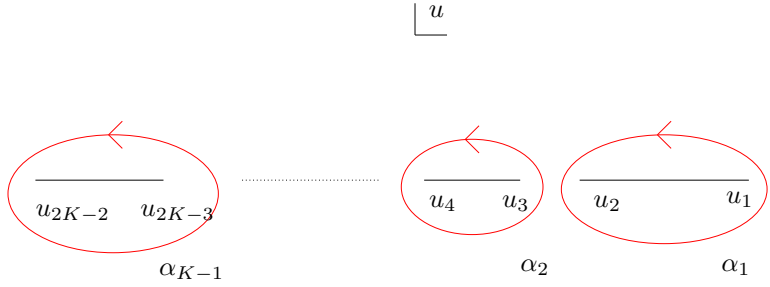}
\caption{The cut $u$-plane corresponding to the curve $\Gamma_{K}$}
\label{Sfig1b}
\end{figure}
\paragraph{}
The Bohr-Sommerfeld conditions 
are expressed in terms of the periods of a certain
meromorphic differential on $\Gamma_{K}$, 
\bea 
\frac{1}{2\pi} \,\oint_{\alpha_{j}}\,u\,\frac{dt}{t} & = &
{l_{j}}\in \mathbb{Z}^{+} \label{BS1} \eea
for $j=1,2,\ldots, K-1$. The positive integers $l_{j}$ which label the states 
are known as filling numbers. As discussed below they correspond to
the total number of Bethe roots associated with the corresponding
cut. Imposing these conditions leads to a
discrete spectrum for the conserved charges, 
\bea {q}_{j} & = &
{q}_{j}\left[l_{1},l_{2},\ldots,l_{K-1}
\right] 
\label{quant} \eea 
and therefore also for the anomalous
dimensions (\ref{gpred}). 
\paragraph{}
The semiclassical limit of large spin can also be studied as a limit
of the Bethe Ansatz which provides the exact solution of the
corresponding quantum spin chain. Exact eigenstates of the spin chain 
are characterised by a set of magnon
rapidities $\{u_{a}\}$, $a=1,2,\ldots,S$ also known as Bethe
roots. These roots solve the Bethe Ansatz equations (BAE), 
\bea 
\left(\frac{u_{a}+i/2}{u_{a}-i/2}\right)^{J} & = & 
\,\,\prod^{S}_{b \neq a}\,\,\left(\frac{u_a-u_b -i}
{u_a-u_b+i}\right) \label{BAE} \eea 
for each $a=1,2,\ldots,S$. The conserved momentum associated with each 
Bethe root is, 
\bea p_{a} & = & \frac{1}{i} \log \left(\frac{u_{a}-\frac{i}{2}}
{u_{a}+\frac{i}{2}}\right) \,\,\,+\,\,\, 2\pi n_{a} 
\eea 
Here $n_a$ is an integer known as the {\em mode number} of the Bethe
root $u_{a}$ which is naturally defined modulo $J$. It is believed
that the BAE (\ref{BAE}) has a unique solution
corresponding to each set of mode numbers. 
Thus a state of the
quantum spin chain is completely specified by its set of mode numbers
$\{n_{a}\}$. Equivalently, the 
state is also characterised by its filling numbers
$l_{j}$ introduced above.  In the context of the Bethe ansatz these 
correspond to the 
number of roots with mode number equal (modulo
$J$) to $j$. Hence they obey $\sum_{j} l_{j} =S$. 
For each solution of the BAE, the formula for the one loop anomalous
dimension is, 
\bea
\Delta-S-J & = & \frac{\lambda}{8\pi^{2}}
\sum_{a=1}^{s}\,\,\frac{1}{u_{a}^{2}+\frac{1}{4}}
\label{exact} \eea
In the large-$S$ limit the roots condense to form cuts in
the complex $u$ plane and the resulting double-cover of the $u$ plane
is precisely the spectral curve $\Gamma_{K}$ introduced above. Here
$K$ corresponds to the number $\leq J$ of filling numbers $l_{j}$
which scale linearly with $S$ as $S\rightarrow \infty$. In this limit
the exact spectrum of one-loop anomalous dimensions 
specified by equations (\ref{BAE},\ref{exact}) goes over to the
semiclassical result specified by (\ref{gpred},\ref{BS1},\ref{quant}).

\section{Classical string theory in $AdS_{3}$}

\subsection{Preliminaries and Conventions}

\subsubsection{Coordinates, string equations of motion and 
Virasoro constraints}

$AdS_3$ space is a 3-dimensional hyperboloid embedded in $\mathbb{R}^{2,2}$ defined by the following constraint:
\begin{equation}
X_\mu X^\mu = -X_0^2-X_1^2+X_2^2+X_3^2 = -1 \nonumber
\label{eq:AdS3constr}
\end{equation}
The $\sigma$-model action, in conformal gauge, is then defined in terms of the embedding coordinates $X_\mu$ as:
\begin{equation}
I = - \frac{\sqrt{\lambda}}{4\pi}\int d\sigma d\tau \left[ G_{\mu\nu} \partial_a X^\mu \partial^a X^\nu + 
 \Lambda \left( X_\mu X^\mu + 1 \right) \right]
\label{eq:AdS3sigmamodelaction}
\end{equation}
where $\lambda=g^{2}N$ is the t'Hooft coupling, $G_{\mu\nu} = \mathrm{diag}(-1,-1,1,1)$ is the $\mathbb{R}^{2,2}$ metric, and the worldsheet indices are contracted with the 2-dimensional Minkowski metric $\eta_{ab} = \mathrm{diag}(-1,1)$.

Once we eliminate the Lagrange multiplier $\Lambda$, the equations of motion for this action become:
\begin{equation}
\partial_+ \partial_- X_\mu - (\partial_+ X^\nu \partial_- X_\nu) X_\mu = 0 \nonumber
\label{eq:AdS3eomemb}
\end{equation}
where we have introduced the light-cone worldsheet coordinates, $\sigma^\pm = (\tau\pm\sigma)/2$. The string must also obey the Virasoro constraints which read:
\begin{equation}
\partial_\pm X^\mu \partial_\pm X_\mu = 0
\label{eq:Virasoropm} \nonumber
\end{equation}

$AdS_3$ can also be parametrised by global coordinates ($t$,$\rho$,$\phi$), which are related to the embedding coordinates as follows:
\begin{eqnarray}
 X_0 & = & \cosh\rho \cos t \nonumber \\
 X_1 & = & \cosh\rho \sin t \nonumber \\
 X_2 & = & \sinh\rho \cos \phi \nonumber \\
 X_3 & = & \sinh\rho \sin \phi \label{eq:AdS3globaltoembcoordtransf} \nonumber
\end{eqnarray}
In terms of these coordinates, the $AdS_3$ line element is:
\begin{equation}
 ds^2 = - \cosh^2\rho \; dt^2 + d\rho^2 + \sinh^2\rho \; d\phi^2
\label{eq:AdS3_line_element_global_coords}
\end{equation}
We also notice that the points $(t,-\rho,\phi)$ and $(t,\rho,\phi+\pi)$ correspond to the same point of the hyperboloid in the embedding space $\mathbb{R}^{2,2}$. Since we want our coordinate chart to be 1-1, we then need to demand that $\rho \geq 0$.

Finally, we introduce the complex coordinates $Z_i$:
\begin{eqnarray}
 Z_1 & = & X_0 + i X_1 = \cosh\rho \; e^{i t} \nonumber \\
 Z_2 & = & X_2 + i X_3 = \sinh\rho \; e^{i \phi}
\label{eq:AdS3defcxcoords}
\end{eqnarray}
In terms of these coordinates, the $AdS_3$ constraint, the equations of motion and the Virasoro constraints are respectively rewritten as:
\begin{eqnarray}
 |Z_1|^2 - |Z_2|^2 & = & 1 \label{eq:AdS3constrcx} \\
 \partial_+\partial_- Z_i - \mathrm{Re} (-\partial_+ Z_1 \partial_- \bar{Z}_1 + \partial_+ Z_2 \partial_- \bar{Z}_2) Z_i & = & 0 \quad i=1,2
   \label{eq:AdS3eomcx} \\
 - \partial_\pm Z_1 \partial_\pm \bar{Z}_1 + \partial_\pm Z_2 \partial_\pm \bar{Z}_2 & = & 0 \label{eq:AdS3Virasorocx}
\end{eqnarray}

\subsubsection{Conserved charges, SU(1,1) and the Monodromy Matrix}
\label{sec:charges_SU11_monodromy}

The action \eqref{eq:AdS3sigmamodelaction} is invariant under global time translations $t \to t + a$, and rotations $\phi \to \phi + b$. By Noether's theorem, the associated conserved charges are the energy $\Delta$ and the angular momentum $S$, given by:
\begin{eqnarray}
 \Delta & = & \frac{\sqrt{\lambda}}{2\pi} \int d\sigma \; \mathrm{Im}(\bar{Z}_1 \partial_\tau Z_1) \label{eq:AdS3_energy} \\
 S & = & \frac{\sqrt{\lambda}}{2\pi} \int d\sigma \; \mathrm{Im}(\bar{Z}_2 \partial_\tau Z_2) \label{eq:AdS3_spin}
\end{eqnarray}
where the integrals are carried out over the entire range of $\sigma$ (e.g. $[0,2\pi]$ for a closed string).

We now recall the fact that it is possible to associate each point in $AdS_3$ with an element $g$ of the group $SU(1,1)$ as follows:
\begin{equation}
 g = \left( \begin{array}{cc}
	Z_1 & Z_2 \\
	\bar{Z}_2 & \bar{Z}_1
     \end{array} \right)
\label{eq:defgSU11elt}
\end{equation}
The matrix $g$ satisfies the $SU(1,1)$ properties:
\begin{eqnarray}
 g^\dag M g & = & M\ , \qquad M = \left( \begin{array}{cc}
                                       1 & 0 \\
                                       0 & -1
                                     \end{array} \right)
\label{eq:SU11property1} \nonumber \\
 \mathrm{det}\: g & = & 1 \label{eq:SU11property2} \nonumber
\end{eqnarray}
by virtue of the $AdS_3$ constraint \eqref{eq:AdS3constrcx}.

The corresponding Lie algebra $\mathfrak{su}(1,1)$ is defined as the space of $2\times 2$ matrices $m$ satisfying:
\begin{equation}
 \mathrm{tr} \; m = 0\:, \qquad m^\dag = - M m M \nonumber
\label{eq:su11properties}
\end{equation}
We choose the set of generators $(s^1, s^2, s^3) = (-i\sigma_3,\sigma_1,-\sigma_2)$ (where $\sigma_1$, $\sigma_2$ and $\sigma_3$ are the Pauli matrices) obeying:
\begin{equation}
 [s^A,s^B] = - 2 \epsilon^{ABC}\eta_{CD}s^D
\label{eq:su11_generators_commutator} \nonumber
\end{equation}
and:
\begin{equation}
 \mathrm{tr} (s^A s^B) = 2 \eta^{AB}
\label{eq:su11_generators_trace}
\end{equation}
where $\eta_{AB}/2$, with $\eta = \mathrm{diag}(-1,1,1)$, is the metric on the Lie algebra.
 
A generic Lie algebra valued quantity $V$ is expressed as:
\begin{equation}
 V = V_A s^A = \frac{1}{2}\eta_{AB}V^A s^B = \frac{1}{2} 
\left( \begin{array}{cc} iV^0 & V^1 + i V^2 \\ V^1 - i V^2 & -iV^0 \end{array} \right)
\label{eq:su11_vector_decomposition}
\end{equation}
In particular we define the $\mathfrak{su}(1,1)$-valued right current $j$:
\begin{equation}
 j_a = g^{-1}\partial_a g= \frac{1}{2}\eta_{AB}j_{a}^A s^B
\label{eq:def_su11_right_current}
\end{equation}
in terms of which the $\sigma$-model action \eqref{eq:AdS3sigmamodelaction} becomes:
\begin{equation}
 I = \frac{\sqrt{\lambda}}{4\pi} \int d^2\sigma \frac{1}{2} \mathrm{tr} (-j_a j^a)
\label{eq:SL2R_principal_chiral_model_action} \nonumber
\end{equation}
which is the action of the  $SL(2,\mathbb{R})$ Principal Chiral model. 

In this formulation the theory is invariant under left and right multiplication by a constant $SU(1,1)$ group element $U_L$/$U_R$:
\begin{equation}
 g \to U_L g\:, \qquad g \to g U_R
\label{eq:SU11_left_and_right_multiplication}
\end{equation}
The associated conserved currents are:
\begin{equation}
 J_L^a = - \frac{\sqrt{\lambda}}{4\pi} l^a\:, \qquad J_R^a = - \frac{\sqrt{\lambda}}{4\pi} j^a
\label{eq:SU11_multiplication_currents} \nonumber
\end{equation}
where $j^a$ is defined in (\ref{eq:def_su11_right_current}) above and 
\begin{equation}
 l_a = (\partial_a g) g^{-1}
\label{eq:eq:def_su11_left_current}
\end{equation}
The equations of motion imply the conservation conditions:
\begin{equation}
 \partial_+ j_- + \partial_- j_+ = -2 \partial^a j_a = 0 \:, \qquad \partial_+ l_- + \partial_- l_+ = -2 \partial^a l_a = 0
\label{eq:j_l_conservation} \nonumber
\end{equation}
and the currents also obey flatness conditions:
\begin{equation}
 \partial_+ j_- - \partial_- j_+ - [j_-,j_+] = 0 \:, \qquad \partial_+ l_- - \partial_- l_+ + [l_-,l_+] = 0
\label{eq:j_l_flatness} \nonumber
\end{equation}
The corresponding conserved charges are:
\begin{equation}
 Q_L = \frac{\sqrt{\lambda}}{4\pi} \int d\sigma l_\tau\:, \qquad Q_R = \frac{\sqrt{\lambda}}{4\pi} \int d\sigma j_\tau
\label{eq:SU11_left_and_right_charges}
\end{equation}
The components of these charges lying in the Cartan subalgebra generated by $s^0$ can be related to the string energy \eqref{eq:AdS3_energy} and $AdS_{3}$ angular momentum \eqref{eq:AdS3_spin}:
\begin{equation}
 Q_L^0 = \Delta - S \:, \qquad Q_R^0 = \Delta + S
\label{eq:SU11_charges_first_component}
\end{equation}

Another important point is the fact that:
\begin{equation}
 -\frac{1}{2} \mathrm{tr}j^2_a = \mathrm{det} j_a = \partial_a Z_1 \partial_a \bar{Z}_1 - \partial_a Z_2 \partial_a \bar{Z}_2
\label{eq:det_and_tr_of_j}
\end{equation}
(where the first equality is just a consequence of the general $\mathfrak{su}(1,1)$ matrix structure \eqref{eq:su11_vector_decomposition}, and $j^2_a$ indicates the matrix square of $j_a$) and thus it is possible to express the Virasoro constraints \eqref{eq:AdS3Virasorocx} in terms of the current:
\begin{equation}
 -\frac{1}{2} \mathrm{tr}j^2_{\pm} = \mathrm{det} \: j_\pm = 0
\label{eq:Virasoro_j}
\end{equation}
Integrability of the string $\sigma$-model follows from the existence of the {\rm Lax connection}:
\begin{eqnarray}
 \mathcal{J}_\tau   (x,\sigma,\tau) & = & \frac{1}{2} \left( \frac{j_+}{1-x} + \frac{j_-}{1+x} \right) \nonumber \\
 \mathcal{J}_\sigma (x,\sigma,\tau) & = & \frac{1}{2} \left( \frac{j_+}{1-x} - \frac{j_-}{1+x} \right)
\label{eq:def_spectral_current} \nonumber
\end{eqnarray}
which is a one-parameter family of $\mathfrak{su}(1,1)$ connections labelled by the spectral parameter $x$. The flatness of the Lax connection, for all $x\in \mathbb{R}$ is equivalent to the string equations of motion given above. This flatness condition in turn implies that the monodromy matrix:
\begin{equation}
 \Omega[x;\sigma_{0},\tau_{0}] = \mathcal{P} \mathrm{exp} \left[
 \int_{\gamma(\sigma_0,\tau_0)} d\sigma^a (-\mathcal{J}_a) \right]
 \,\,\,\, \in \,\,SL(2,\mathbb{R})
\label{eq:def_monodromy_matrix}
\end{equation}
evolves by conjugation in the group. Here $\mathcal{P} \mathrm{exp}$ indicates the path-ordered exponential and the integral is calculated along an arbitrary closed path $\gamma$ with base point $(\sigma_{0},\tau_{0})$. In the following we specialize to a path which winds once around the string at constant worldsheet time $\tau=\tau_{0}$. In this case 
\bea 
\Omega\left[x;\tau \right] & = &
\,\,\mathcal{P}\,\,\exp\left[\frac{1}{2}
\int_{0}^{2\pi}\,d\sigma \,\, \left( \frac{j_{+}}{x-1} \,\,+\,\,
\frac{j_{-}}{x+1}\right )\right] \,\,\,\,\in\,\,\,\, SL(2,\mathbb{R})
\label{eq:monodromy_matrix_for_large_spikes} \eea 
and the corresponding eigenvalues $t_{\pm}=\exp(\pm i\,p(x))$ are
$\tau$-independent for all values of the spectral parameter $x$. It is
convenient to consider the analytic continuation of the monodromy
matrix $\Omega[x;\tau]$ and of the quasi-momentum $p(x)$ to complex
values of $x$. In this case $\Omega$ will take values in
$SL(2,\mathbb{C})$ and appropriate reality conditions must be imposed
to recover the physical case.   
\paragraph{}
The eigenvalues $t_{\pm}(x)$ are two branches of an analytic
function defined on the spectral curve, 
\bea 
\Sigma_{\Omega}\,\,: \qquad{} \quad{} 
t \, + \, \frac{1}{t} \,\, = \,\, \mathrm{tr} \: \Omega[x;\tau] \,\, = \,\, 2 \cos p(x) 
\qquad{} t, \,\,x\in \mathbb{C} \label{stringcurve} \eea 
This curve corresponds to a double cover of the complex $x$-plane with
branch points at the simple zeros of the discriminant
$D=4\sin^{2}p(x)$. In the following sections we will determine this 
curve for a variety of solutions in the limit of large angular
momentum.  

\subsection{The $N$-folded string} 

\subsubsection{General Properties}
\label{general_properties_GKP_Nfolded}

The $N$-folded GKP string is a simple generalisation of the solution that was first described in \cite{GKP}. We start with the following ansatz\footnote{In the following we reserve the notation $\tau$ and $\sigma$ for worldsheet coordinates with periodicity $\sigma +2\pi$. These are related to the present worldsheet coordinates $\tilde{\tau}$ and $\tilde{\sigma}$ by a rescaling we will describe below.}: $t = \tilde{\tau}$, $\rho = \rho(\tilde{\sigma})$, $\phi = \phi_0 + \omega \tilde{\tau}$. The string equations of motion in conformal gauge then reduce to:
\begin{equation}
 (\partial_{\tilde{\sigma}} \rho)^2 = \cosh^2 \rho - \omega^2 \sinh^2 \rho
\label{eq:EOM_conf_gauge_for_GKP_ansatz}
\end{equation}
which we can integrate as:
\begin{equation}
 \tilde{\sigma} = \int_0^{\rho(\tilde{\sigma})} \frac{dy}{\sqrt{\cosh^2 y - \omega^2 \sinh^2 y}}
\label{eq:EOM_integrated_conf_gauge_for_GKP_ansatz}
\end{equation}
with an appropriate choice of sign for the square root in the integrand. As we want the integrand to be real, we have to restrict $\rho$ to the
range $\rho(\tilde{\sigma}) \in [0,\rho_1]$, where $\coth{\rho_1} = \omega$ with $\omega > 1$. In the limit $\omega\to 1$, we have $\rho_1 \to + \infty$ and the bound is lifted. The solution to \eqref{eq:EOM_integrated_conf_gauge_for_GKP_ansatz} is:
\begin{equation}
 \rho(\tilde{\sigma}) = - i \: \mathrm{am} (i \tilde{\sigma}|\sqrt{1-\omega^2})
\label{eq:GKP_rho_of_sigmatilde}
\end{equation}
which yields:
\begin{eqnarray}
  \coth\rho(\tilde{\sigma}) & = & \frac{\omega}{\mathrm{sn} \left( \omega\tilde{\sigma} \left| \frac{1}{\omega} \right) \right.}
     \label{eq:coth_rho_GKP_preliminary}
\end{eqnarray}
This expression is periodic in $\tilde{\sigma}$ with period $(4/\omega)\mathbb{K}(1/\omega) \equiv 4\tilde{L}$. However, to impose the condition $\rho\geq0$ we must restrict the solution to the first half-period $\tilde{\sigma}\in[0,2\tilde{L}]$. In this interval $\rho$ increases from zero to its maximum value $\rho_1$ at $\tilde{\sigma}=\tilde{L}$ and then returns to zero at $\tilde{\sigma}=2\tilde{L}$. The snapshot of the string at fixed global time $t$ consists of two straight segments of string connecting the origin of the global coordinates with the point $(\rho_1,\phi(t))$. We can add a second pair of segments, stretching out in the opposite direction, by gluing a similar solution, with $\tilde{\sigma}$ replaced by $\tilde{\sigma} - 2 \tilde{L}$ (translations of $\tilde{\sigma}$ are clearly a symmetry of the differential equation \eqref{eq:EOM_conf_gauge_for_GKP_ansatz}) and $\phi(t)$ replaced by $\phi(t)+\pi$ (this only amounts to a change in the value of $\phi_0$, which is arbitrary). In this way we obtain the composite solution, 
\begin{eqnarray}
 \rho = \rho(\tilde{\sigma}) & \phi = \phi_0 + \omega \tilde{\tau} & \mathrm{for} \: \tilde{\sigma} \in [0,2\tilde{L}] \nonumber\\
 \rho = \rho(\tilde{\sigma}-2\tilde{L}) & \phi = \phi_0 + \omega \tilde{\tau} + \pi & \mathrm{for} \: \tilde{\sigma} \in [2\tilde{L},4\tilde{L}]
\label{eq:original_GKP_solution} \nonumber
\end{eqnarray}
where now $\tilde{\sigma}$ runs over a full period $4\tilde{L}$ but, because of the gluing, we preserve the condition $\rho\geq 0$. The solution corresponds to a folded, closed string rotating about its midpoint which lies at the origin. The string has two cusps at the points $\tilde{\sigma}=\tilde{L}$ and $\tilde{\sigma}=3\tilde{L}$.
\paragraph{}
The above solution can be easily generalised by allowing $\tilde{\sigma}$ to range over N periods\footnote{Here $[X]$ denotes the greatest integer less than $X$.}:
\begin{eqnarray}
 \rho & = & \rho \left( \tilde{\sigma}- l(\tilde{\sigma}) (2\tilde{L})
     \right) \quad \textrm{where}\,\,\, l(\tilde{\sigma}) = \left[
     \frac{\tilde{\sigma}}{2\tilde{L}} \right]\nonumber\\
 \phi & = & \phi_1 + \omega \tilde{\tau} = \begin{cases}
              \phi_0 + \omega \tilde{\tau} & \text{if } l(\tilde{\sigma}) \textrm{ is even} \\
              \phi_0 + \omega \tilde{\tau}+\pi & \text{if } l(\tilde{\sigma}) \textrm{ is odd}
            \end{cases}
\label{eq:GKP_Nfolded_solution}
\end{eqnarray}
were now $\tilde{\sigma}\in[0,4N \tilde{L}]$. The image of the string in the target space is unchanged, but the increased range of the coordinate $\tilde{\sigma}$ corresponds to a string folded N times upon itself. We define:
\begin{equation}
 L = 4 N \tilde{L} = \frac{4 N}{\omega} \mathbb{K}
\label{eq:def_L_max_value_of_sigmatilde_GKP_Nfolded} \nonumber
\end{equation}
as the upper extremum of the range of $\tilde{\sigma}$, and:
\begin{equation}
 \tilde{\sigma}_* = \tilde{\sigma}- l(\tilde{\sigma}) (2\tilde{L})
\label{eq:def_sigmatilde_star} \nonumber
\end{equation}
where we have introduced the shorthand notation $\mathbb{E}\equiv \mathbb{E}(1/\omega)$ and $\mathbb{K}\equiv \mathbb{K}(1/\omega)$, which we will use throughout the rest of this section for the elliptic integrals of the first and second kind.
\paragraph{}
We note that this solution has $K=2N$ cusps, that is two for each fold of the string, located at the tips $\rho=\rho_1$ of the line segment, which we can identify with the following worldsheet positions:
\begin{equation}
 \tilde{\sigma}_m = (2m+1) \tilde{L} \:, \quad m = 0,\ldots,K-1
\label{eq:cusp_positions_Nfolded_GKP_sigmatilde}
\end{equation}
Cusps are points along the string where the unit normalised tangent vector has a discontinuity. The only way in which this can happen without compromising the smoothness of the worldsheet is that all components of the tangent vector vanish at the point, so that its direction is actually allowed to change discontinuously even though the vector itself varies smoothly. Therefore, spikes are points at which the derivatives with respect to $\tilde{\sigma}$ of all target space coordinates vanish. Here, we can easily see that $\partial_{\tilde{\sigma}} Z_1 = 0 = \partial_{\tilde{\sigma}} Z_2$ at $\tilde{\sigma} = \tilde{\sigma}_m$, for all $m$, as required. As $\omega\to 1$, $\rho_1$ tends to $+\infty$, and hence the spikes touch the boundary of $AdS_3$ and the string becomes infinitely long.

\subsubsection{Energy, angular momentum and large angular momentum behaviour}
\label{sec:Nfolded_GKP_E_S_large_S_behaviour}

The energy and the angular momentum of the solution \eqref{eq:GKP_Nfolded_solution} can straightforwardly be computed from \eqref{eq:AdS3_energy} and \eqref{eq:AdS3_spin}:
\begin{eqnarray}
 \Delta & = & 2N \frac{\sqrt{\lambda}}{2\pi} \int_0^{2 \tilde{L}} 
d\tilde{\sigma} \frac{1}{\mathrm{dn}^2 \left(\omega \tilde{\sigma}_* \left|
    \frac{1}{\omega} \right) \right.} = K \frac{\sqrt{\lambda}}{\pi} \frac{\omega}{\omega^2 - 1} \mathbb{E} \label{eq:E_Nfolded_GKP} \nonumber\\
 S & = & 2N \frac{\sqrt{\lambda}}{2\pi} \frac{1}{\omega} 
\int_0^{2 \tilde{L}} d\tilde{\sigma} \frac{\mathrm{sn}^2 \left(\omega \tilde{\sigma}_* \left|
    \frac{1}{\omega} \right) \right.}{\mathrm{dn}^2 \left(\omega
    \tilde{\sigma}_* \left| \frac{1}{\omega} \right) \right.} 
\nonumber \\
   & = & K \frac{\sqrt{\lambda}}{\pi} \left[ \frac{\omega^2}{\omega^2 - 1} \mathbb{E} - \mathbb{K} \right] \label{eq:S_Nfolded_GKP} \nonumber
\end{eqnarray}
As we could expect due to the periodicity of the integrands, the values are just N times the original GKP values, as they appear in \cite{JJ}.

We are interested in the large $S$ behaviour of this solution, which corresponds to the limit $\omega\to 1$. Both $\Delta$ and $S$ diverge in this limit and, if we define:
\begin{equation}
 \omega = 1 + \eta
\label{eq:def_eta_omega_goes_to_1} \nonumber
\end{equation}
then their respective behaviours are given by:
\begin{eqnarray}
 \Delta & = & K \frac{\sqrt{\lambda}}{2\pi \eta} - K \frac{11 \sqrt{\lambda}}{32\pi} \log\eta - K \frac{ \sqrt{\lambda}}{64\pi} [13 -
          44 \log (2\sqrt{2})] + O(\eta \log \eta) \label{eq:E_behaviour_as_omega_goes_to_1_Nfolded_GKP} \\
 S & = & K \frac{\sqrt{\lambda}}{2\pi \eta} + K \frac{5 \sqrt{\lambda}}{32\pi} \log\eta + K \frac{ \sqrt{\lambda}}{64\pi} [19 -
          20 \log (2\sqrt{2})] + O(\eta \log \eta) \label{eq:S_behaviour_as_omega_goes_to_1_Nfolded_GKP}
\end{eqnarray} 
Thus each spike contributes $\sqrt{\lambda}/(2\pi\eta)$ to the leading order term. 
\paragraph{}
From these equations, we can deduce the leading behaviour of the anomalous dimension:
\begin{equation}
 \Delta - S = \frac{K\sqrt{\lambda}}{2\pi} \log \left( \frac{2\pi
 S}{K\sqrt{\lambda}} \right) + \frac{K\sqrt{\lambda}}{2\pi} (3\log 2-1) + O(\eta\log\eta)
\label{eq:E_S_omega_to_1_behaviour}
\end{equation}
which exhibits the same logarithmic growth found in gauge theory.

\subsubsection{Spectral curve for large S}
\label{sec:spectral_curve_Nfolded_GKP}

In this section we will compute the monodromy matrix and spectral curve of the $N$ folded string solution discussed above. As a starting point, we need to compute the time-like component of the right current $j$. Using the ansatz $\rho = \rho(\tilde{\sigma})$ we obtain:
\begin{eqnarray}
 j^0_{\tilde{\tau}} & = & 2 [(\partial_{\tilde{\tau}} t) \cosh^2\rho + (\partial_{\tilde{\tau}} \phi) \sinh^2\rho] \nonumber\\
 j^1_{\tilde{\tau}} + i j^2_{\tilde{\tau}} & = & 2 i \sinh\rho \cosh\rho \: e^{i(\phi-t)} [(\partial_{\tilde{\tau}} t) +
  (\partial_{\tilde{\tau}} \phi)]
\label{eq:components_of_j_tau_in_terms_of_t_phi_rho}
\end{eqnarray}
Evaluating this on the $N$-folded string solution we obtain, 
\begin{eqnarray}
 j^0_\tau (\tau,\sigma) & = & \frac{2 K \mathbb{K}}{\pi\omega} \frac{\left[ 1 + \frac{1}{\omega} \mathrm{sn}^2 \left( \omega
        \sigma_* \left| \frac{1}{\omega} \right) \right] \right.}{\mathrm{dn}^2 \left( \omega \sigma_* \left| \frac{1}{\omega} \right) \right.}
         \nonumber \\
 j^1_\tau (\tau,\sigma) + i j^2_\tau (\tau,\sigma) & = & i
        \frac{2 K \mathbb{K}}{\pi} \frac{\omega + 1}{\omega^2}
        \frac{\mathrm{sn} 
\left(\omega \sigma_* \left| \frac{1}{\omega} \right) \right.}{\mathrm{dn}^2 \left( \omega \sigma_* \left| \frac{1}{\omega}
          \right) \right.} e^{i \left[ \phi_1 + (\omega - 1) \frac{K \mathbb{K}}{\pi\omega} \tau \right]}
\label{eq:j_tau_in_terms_of_tau_sigma_Nfolded_GKP}
\end{eqnarray}
Here we have introduced rescaled worldsheet coordinates $(\tau,\sigma)$ with $\sigma\in [0,2\pi]$:
\begin{equation}
 (\tau,\sigma) = \frac{2\pi}{L} (\tilde{\tau},\tilde{\sigma}) =  \frac{\pi\omega}{K \mathbb{K}}(\tilde{\tau},\tilde{\sigma}) 
\label{eq:def_rescaled_coords_Nfolded_GKP} \nonumber
\end{equation}
and $\sigma_*$ is just $\tilde{\sigma}_*$ written as a function of $\sigma$. We can also express the worldsheet positions of the cusps 
\eqref{eq:cusp_positions_Nfolded_GKP_sigmatilde} in terms of the rescaled coordinate $\sigma$ as:
\begin{equation}
 \sigma_m = (2m+1) \frac{\pi}{K} \:, \qquad m=0,\ldots,K-1
\label{eq:cusp_positions_Nfolded_GKP_sigma} \nonumber
\end{equation}
\paragraph{}
The next step is to take the limit $\omega\rightarrow 1$ so that the angular momentum $S$ of the solution diverges. The key point here is that, as $S\rightarrow \infty$, the charge density is dominated by the vicinity of the cusp points $\sigma=\sigma_{m}$. To demonstrate this we expand around the $m$-th cusp point setting:
\begin{equation}
 \sigma = \sigma_m + \hat{\sigma} \:, \qquad \textrm{with } |\hat{\sigma}| < \frac{\pi}{K}
\label{eq:def_sigma_near_cusps_Nfolded_GKP}
\end{equation}
which is equivalent to  $\sigma_* = \tilde{L} + \hat{\sigma}
K \mathbb{K}/ (\pi\omega)$ for each $m$. 
We can then use the quarter-period transformation formulae for the elliptic functions appearing in \eqref{eq:j_tau_in_terms_of_tau_sigma_Nfolded_GKP} to get:
\begin{eqnarray}
 \mathrm{sn} \left( \omega \sigma_* \left| \frac{1}{\omega} \right)
 \right. & = & \mathrm{sn} \left( \mathbb{K} + 
\frac{K \mathbb{K}}{\pi}
  \hat{\sigma} \left| \frac{1}{\omega} \right) \right. = 
\mathrm{cd} \left( \frac{K \mathbb{K}}{\pi} \hat{\sigma} \left| \frac{1}{\omega} \right)
   \right. \nonumber\\
 \mathrm{dn} \left( \omega \sigma_* \left| \frac{1}{\omega} \right)
  \right. & = & \mathrm{dn} \left( \mathbb{K} + \frac{K \mathbb{K}}
{\pi}
  \hat{\sigma} \left| \frac{1}{\omega} \right) \right. = \sqrt{1
 -\frac{1}{\omega^2}} \; \mathrm{nd} \left( \frac{K \mathbb{K}}{\pi} 
\hat{\sigma}
   \left| \frac{1}{\omega} \right) \right.
\label{eq:half_period_transf_for_sn_dn_Nfolded_GKP} \nonumber
\end{eqnarray}
As we are interested in the limit $\omega\rightarrow 1$ we can 
use the standard series expansions for $\mathrm{cd}(z|k)$ and
$\mathrm{nd}(z|k)$ in powers of $(k-1)$ \cite{BF}, with   
\bea 
z &= & \frac{K \mathbb{K}}{\pi} 
\hat{\sigma}
\nn \eea 
and $k=1/\omega$. Note however that 
$\mathbb{K}\simeq - (1/2) \ln (\omega-1)$ for 
$\omega\rightarrow 1$. This implies we must consider a limit where 
not only $k\to 1$, but also $z\to\pm\infty$, depending on the sign of
$\hat{\sigma}$. However, since $|z|=K \mathbb{K}|\hat{\sigma}|/\pi <
\mathbb{K}$, i.e. $z$ is always within the first quarter-period in
both directions, one can check that higher order terms remain
suppressed. We thus consider only the lowest order terms in these series,
which give the leading behaviour of 
$j_{\tau}$ near each spike as:
\begin{eqnarray}
 j_\tau^0(\tau,\sigma) & \simeq & \frac{2 K \mathbb{K}}{\pi}\frac{1}{\eta \cosh^2 \left( \frac{K \mathbb{K}}{\pi} \hat{\sigma} \right)} \nonumber
  \\
 j_\tau^1(\tau,\sigma) + i j_\tau^2(\tau,\sigma) & \simeq & (-1)^m i \frac{2 K \mathbb{K}}{\pi} \frac{e^{i\phi_0}}{\eta \cosh^2 \left(
  \frac{K \mathbb{K}}{\pi} \hat{\sigma} \right)}
\label{eq:leading_behaviour_for_cpts_of_j_tau_GKP_Nfolded}
\end{eqnarray}
Here the factor $(-1)^m$ comes from the extra $\pi$ which is added to $\phi$ every other period, as specified in \eqref{eq:GKP_Nfolded_solution}, and consequently affects the contribution of every other cusp. 
\paragraph{}
As the constant $\mathbb{K}$ diverges like $\log \eta$ for $\omega\to 1$, the resulting expression for $j_{\tau}$ is singular in this limit. To obtain a finite result, we first define the normalised charge density:
\begin{equation}
 \mu^A(\tau,\sigma) = \lim_{\omega\to 1} \frac{\sqrt{\lambda}}{8\pi S} j^A_\tau (\tau,\sigma)
\label{eq:def_normalised_charge_density}
\end{equation}
which satisfies:
\begin{equation}
 \int_0^{2\pi} \; d\sigma \vec{\mu} (\tau,\sigma) = \begin{pmatrix}
	                                                 1 \\
	                                                 0 \\
	                                                 0
 				                                         
\end{pmatrix}
\label{eq:property_of_renorm_charge_density}
\end{equation}
on highest weight states. Using \eqref{eq:leading_behaviour_for_cpts_of_j_tau_GKP_Nfolded} and \eqref{eq:S_behaviour_as_omega_goes_to_1_Nfolded_GKP}, we can compute the contribution to the normalised charge density from the region near the $m$-th cusp and then sum it over $m=0,\ldots,K-1$ to get:
\begin{eqnarray}
 \mu^0(\tau,\sigma) & = & \frac{1}{K} \sum_{m=0}^{K-1} \delta(\sigma - \sigma_m) \nonumber\\
 \mu^1(\tau,\sigma) + i \mu^2(\tau,\sigma) & = & \frac{e^{i\left( \phi_0 + \frac{\pi}{2}\right)}}{K} \sum_{m=0}^{K-1} (-1)^m \delta(\sigma -
  \sigma_m)
\label{eq:cpts_of_mu_GKP_Nfolded} \nonumber
\end{eqnarray}
where we have used the identity:
\begin{equation}
 \lim_{\epsilon\to 0} \frac{1}{2 \epsilon} \frac{1}{\cosh^2 \left( \frac{x}{\epsilon} \right)} = \delta(x)
\label{eq:identity_for_delta_function}
\end{equation}
and we have eliminated $\hat{\sigma}$ in favour of $\sigma$ according to \eqref{eq:def_sigma_near_cusps_Nfolded_GKP}. It is now easy to see
that the normalisation condition \eqref{eq:property_of_renorm_charge_density} is indeed satisfied. Finally, using \eqref{eq:def_normalised_charge_density}, we can obtain the large-$S$ asymptotics of the current $j_{\tau}$ in the form:
\begin{eqnarray}
j^A_\tau (\tau,\sigma)  & \rightarrow  & \frac{8\pi}{\sqrt{\lambda}} \sum_{m=0}^{K-1} L_m^A \delta(\sigma - \sigma_m)
\label{eq:def_spin_vector_at_each_cusp}
\end{eqnarray}
with:
\begin{equation}
 \vec{L}_m = \frac{S}{K} \begin{pmatrix}
	                        1 \\
	                        (-1)^{m+1} \sin\phi_0 \\
	                        (-1)^m \cos\phi_0
                         \end{pmatrix} \qquad m=0,\ldots,K-1
\label{eq:spin_vector_at_mth_cusp_Nfolded_GKP}
\end{equation}
\paragraph{}
The $\mathfrak{su}(1,1)$-valued quantities $L_m$ played an important 
role in the proposal of \cite{D1}. In particular, 
for generic large-$S$ solutions, they can be mapped onto the classical
spins of the dual gauge theory spin chain. 
For this particular solution, we can easily verify the properties:
\begin{equation}
 \sum_{m=0}^{K-1} \vec{L}_m = \begin{pmatrix}
	                         S \\
	                         0 \\
	                         0
                          \end{pmatrix} \:, \qquad \eta_{AB} L^A_m L^B_m = 0 \quad \textrm{for } m=0,\ldots,K-1
\label{eq:properties_of_spin_vectors}
\end{equation}
which are significant in this context as the string theory
counterparts of the relations (\ref{Cas}) and (\ref{hw}). The first equality is just a
rephrasing of the normalisation property
\eqref{eq:property_of_renorm_charge_density}, while the second can be
seen as a consequence of the Virasoro constraints
\eqref{eq:Virasoro_j}. In particular, as $j_\sigma(\tau,\sigma_m)=0$,
$\forall m$, we have $j_\pm (\tau,\sigma_m) = j_\tau (\tau,\sigma_m)$,
and the 
Virasoro constraints at the spikes become:
\begin{equation}
 - \frac{1}{2} \lim_{\sigma \to \sigma_m} 
\mathrm{tr} [ j^2_\pm (\tau,\sigma) ] = - \frac{1}{2} \lim_{\sigma \to \sigma_m} \mathrm{tr}
  [ j^2_\tau (\tau,\sigma) ] = 0
\label{eq:Virasoro_near_spikes} \nonumber
\end{equation}
If we now also take the limit $\omega\to 1$, substitute in equation \eqref{eq:def_spin_vector_at_each_cusp}, decompose $j_\tau$ onto the generators $s^A$ as in \eqref{eq:su11_vector_decomposition} and use property \eqref{eq:su11_generators_trace}, we see that the only way in which this condition can hold is that the spin vectors satisfy the second equation \eqref{eq:properties_of_spin_vectors}.
\paragraph{}
Our next goal is to compute the limiting form of the monodromy matrix (\ref{eq:monodromy_matrix_for_large_spikes}) in the large spin limit. This calculation was described for the $N=1$ case in \cite{D1} and we will follow the same steps here. To keep the exponent of the monodromy matrix
\eqref{eq:monodromy_matrix_for_large_spikes} finite as $S\rightarrow\infty$  we are forced to scale the spectral parameter as $x \sim S$. The limiting form of the monodromy matrix then becomes:
\begin{equation}
 \Omega[x;\tau] \simeq \mathcal{P} \mathrm{exp} \left[ \frac{1}{x} \int_0^{2\pi} d\sigma j_\tau (\tau,\sigma) \right]
\label{eq:approx_monodromy_matrix_as_omega_to1_1} \nonumber
\end{equation}
We can now replace $j_\tau$ by its limiting form \eqref{eq:def_spin_vector_at_each_cusp}. The resulting sum of $\delta$-functions in the integrand converts the path-ordered exponential into a finite ordered product of exponentials: 
\begin{equation}
 \Omega[x;\tau] \simeq \prod_{m=0}^{K-1} \mathrm{exp} \left[ \frac{4\pi}{\sqrt{\lambda}} \frac{1}{x} L^A_m s^B \eta_{AB} \right]
\label{eq:approx_monodromy_matrix_as_omega_to1_2}
\end{equation}
where we have also expressed $L_m$ in terms of the $\mathfrak{su}(1,1)$ generators, according to \eqref{eq:su11_vector_decomposition}. We then observe that:
\begin{equation}
 (\eta_{AB} L^A_m s^B)^2 = \frac{1}{2} \mathbb{I} \: \eta_{AB} L^A_m L^B_m = 0
\label{eq:square_of_matrix_in_exp_vanishes_Nfolded_GKP_monodromy} \nonumber
\end{equation}
as a consequence of the fact that the generators $s^A = (-i\sigma_3,\sigma_1,-\sigma_2)^A$ satisfy:
\begin{equation}
 \{ s^A, s^B \} = 2 \eta^{AB}
\label{eq:anticommutator_of_our_su11_generators} \nonumber
\end{equation}
and of the second property \eqref{eq:properties_of_spin_vectors} of the spin vectors. Therefore, the series expansion for the exponential in \eqref{eq:approx_monodromy_matrix_as_omega_to1_2} actually truncates at the linear term:
\begin{equation}
 \Omega[x;\tau] \simeq \prod_{m=0}^{K-1} \left[ \mathbb{I} + \frac{1}{u} \eta_{AB} L^A_m s^B \right]
\label{eq:approx_monodromy_matrix_as_omega_to1_3} \nonumber
\end{equation}
where we have defined a rescaled spectral parameter $u = x \sqrt{\lambda} / (4\pi)$. Using the explicit form \eqref{eq:spin_vector_at_mth_cusp_Nfolded_GKP} for the spin vectors $L_m$ we obtain: 
\begin{equation}
 \Omega[x;\tau] \simeq \frac{1}{u^n} \prod_{m=0}^{K-1} \mathbb{L}_m (u)
\label{eq:monodromy_matrix_as_omega_to1_utilde}
\end{equation}
where:
\begin{equation}
 \mathbb{L}_m (u) = \begin{pmatrix}
	                       u + \frac{iS}{K}                    & (-1)^m \frac{iS}{K} e^{i\phi_0} \\
	                       -(-1)^m \frac{iS}{K} e^{-i\phi_0}   & u - \frac{iS}{K}
                        \end{pmatrix}
\label{eq:L_m(utilde)_for_Nfolded_GKP}
\end{equation}
As explained in \cite{D1}, the matrices $\mathbb{L}_{m}$ 
are the string theory analogues of the 
Lax matrices (\ref{lax}) of the gauge theory spin chain. 
In the present case, 
it is easily seen that the matrices $\mathbb{L}_m (u)$ only depend on the parity of $m$. Therefore, if we define $\mathbb{L} (u) = \mathbb{L}_0 (u) \mathbb{L}_1 (u)$, we can write the monodromy matrix as:
\begin{equation}
 \Omega[x;\tau] \simeq \frac{1}{u^n} [\mathbb{L} (u)]^{\frac{K}{2}}
\label{eq:monodromy_matrix_for_Nfolded_GKP} \nonumber
\end{equation}
It follows that the eigenvalues of $\Omega$ can be expressed in terms of the eigenvalues of the matrix $\mathbb{L} (u)$, which can be
evaluated explicitly as:
\begin{equation}
 \kappa_{\pm} = u^2 - \frac{2 S^2}{K^2} \pm 2 \sqrt{\frac{S^4}{K^4} - \frac{u^2 S^2}{K^2}}
\label{eq:evalues_of_L(utilde)_Nfolded_GKP} \nonumber
\end{equation}
Finally we can write the trace of the monodromy matrix as:
\begin{eqnarray}
 \mathrm{tr} \: \Omega [x] & = & \frac{\kappa_+^\frac{K}{2}+\kappa_-^\frac{K}{2}}{u^K} \nonumber \\
                             & = & \left( 1 - \frac{2 S^2}{K^2 u^2} + i \sqrt{\frac{4 S^2}{K^2 u^2} - \frac{4 S^4}{K^4 u^4}}
                              			\right)^\frac{K}{2} + \left( 1 - \frac{2 S^2}{K^2 u^2} - i \sqrt{\frac{4 S^2}{K^2 u^2} -
                              			 \frac{4 S^4}{K^4 u^4}} \right)^\frac{K}{2} \nonumber \\
                             & = & 2 T_\frac{K}{2} \left( 1 - \frac{2 S^2}{K^2 u^2} \right) = 2 T_K \left( \sqrt{1 - \frac{S^2}{K^2 u^2}}
                              \right) = 
2 \cos \left[ K \sin^{-1} \left( \frac{S}{K u} \right) \right]
\label{eq:tr_Omega_Nfolded_GKP}
\end{eqnarray}
where $\mathrm{T}_k (y)$ is the Chebyshev polynomial of the first kind:
\begin{equation}
 T_k (y) = \frac{1}{2} \left[ \left( y + i \sqrt{1-y^2} \right)^k + \left( y - i \sqrt{1-y^2} \right)^k \right] = \cos (k \arccos y)
\label{eq:def_Tk(y)_Chebyshev}
\end{equation}
This is a polynomial in $y$ of degree $k$. We recall from the discussion of the monodromy matrix at the end of section \ref{sec:charges_SU11_monodromy}, that its eigenvalues are $t_\pm = \exp (\pm i p(u))$, so that:
\begin{equation}
 \mathrm{tr} \: \Omega [x] = 2 \cos p(u)
\label{eq:trace_of_monodromy_matrix_in_terms_of_quasi-momentum} \nonumber
\end{equation}
which, together with \eqref{eq:tr_Omega_Nfolded_GKP}, yields the following expression for the quasi-momentum associated with the N-folded GKP solution:
\begin{equation}
 p(u) = K \sin^{-1} \left( \frac{S}{K u} \right)
\label{eq:quasi-momentum_N-folded_GKP}
\end{equation}
\paragraph{}
In summary, the string theory spectral curve (\ref{stringcurve}) for the $N$-folded GKP solution takes the explicit form:
\bea 
\Sigma_{\Omega}\,\,: \qquad{} \qquad{} t\,\,+\,\,\frac{1}{t} & = & 
2 \cos \left[ K \sin^{-1} \left( \frac{S}{K u} \right)\right] \label{curve2}
  \eea 
We may now compare this directly with the gauge theory spectral curve $\Gamma_{K}$ which can be written in the form: 
\bea 
\Gamma_{K} \,:\qquad{} \qquad{} t\,\, + \,\,\frac{1}{t} & = & \hat{\mathbb{P}}_{K}
\left(\frac{1}{{u}}\right) =  2\,\,+\,\,\frac{q_{2}}
{{u}^{2}}\,\,+\,\, \frac{q_{3}}
{{u}^{3}}\,\,+\,\,\ldots\,\,+\,\, \frac{q_{K}}
{{u}^{K}} \label{eq:general_curve_from_spin_chain} \eea
We then find that the limiting string theory curve
corresponds to a particular point in the moduli space of the gauge
theory curve where the conserved charges $q_{k}$ take the 
particular values:
\begin{equation}
 q_k = \left( \frac{2 S}{K} \right)^k \sum_{1\leq j_1 < j_2 < \ldots < j_k \leq K} \prod_{l=1}^k \sin \left[ \frac{\pi}{2}
  (j_{l+1} - j_l) \right] \:, \quad k=2, 4, \ldots, K
\label{eq:qktilde_from_spin_chain_N-folded_GKP}
\end{equation}
where $j_{k+1} \equiv j_1$ and we notice that $q_k = 0$ for odd $k$ (the normalisation $q_2 = - S^2$ is checked in appendix \ref{sec:app_computing_qtilde_2_for_N-folded_GKP_and_K}).
\paragraph{}
To characterise more precisely the curve corresponding to the $N$-folded string solution it is useful to determine the pattern of the branch points of the spectral curve (\ref{curve2}), which coincide with the simple zeros of the discriminant $D = 4 \sin^2 p(u)$:
\begin{equation}
 D \left(u \right) = 4 \left[ 1 - T_\frac{K}{2}^2 \left( 1 - \frac{2 S^2}{K^2 u^2} \right) \right] = 
  4 \left( \frac{4 S^2}{K^2 u^2} - \frac{4 S^4}{K^4 u^4} \right) U_{\frac{K}{2}-1}^2 \left( 1 - \frac{2 S^2}{K^2 u^2} \right)
\label{eq:rewriting_Q(1/utilde)_GKP_Nfolded_1}
\end{equation}
where $U_k (y)$ is the Chebyshev polynomial of the second kind:
\begin{eqnarray}
 U_k (y) & = & \frac{1}{2i\sqrt{1-y^2}} \left[ \left( y + i
   \sqrt{1-y^2} \right)^{k+1} - 
\left( y - i \sqrt{1-y^2} \right)^{k+1} \right]
  \nonumber \\
         & = & \frac{\sin[(k+1)\arccos y]}{\sin \arccos y}
\label{eq:def_Uk(y)_Chebyshev} \nonumber
\end{eqnarray}
The zeros of the discriminant can then be determined as:
\begin{equation}
 \begin{array}{lc}
  u^{-1}  =  \pm \frac{K}{S}  & \qquad \textrm{simple} \\
  u^{-1}  =  0      & \qquad \textrm{double} \\
  u^{-1} = \pm \frac{K}{S} \sin \left( \frac{k \pi}{K} \right) \:, \qquad k=1,\ldots,\frac{K}{2}-1 & \qquad
   \textrm{double}
 \end{array}
\label{eq:pattern_of_zeros_of_Q_n_for_Nfolded_GKP}
\end{equation}
which all lie in the interval $u^{-1} \in [-K/S,K/S]$. The presence of $K-1$ double zeros indicates the degeneration of the spectral curve to genus zero. In fact, the Abelian integral $p(u)$ from \eqref{eq:quasi-momentum_N-folded_GKP} is analytic on the complex plane, except for the following singularities: a logarithmic branch point at $u = 0$ and two square root branch points at $u = \pm S/K$. We can make $p(u)$ single-valued by introducing a single branch cut connecting these three points along the real axis. Correspondingly, the differential $dp(u)$ is given by:
\begin{equation}
 dp(u) = - \frac{S \: d u}{ u \sqrt{u^2 - \frac{S^2}{K^2}}}
\label{eq:dp(x)_N-folded_GKP} \nonumber
\end{equation}
and it has a simple pole at $u=0$, two square root branch points at $u = \pm S/K$ and the corresponding two square root branch points at infinity. It can be made single-valued by introducing the same cut along the real axis. This single cut defines a Riemann surface of genus zero. 
\paragraph{}
From the gauge theory point of view, as explained in \cite{BGK2}, the
same degenerate curve arises as a limiting case of a two-cut solution,
where the two cuts collide and merge into one at the origin. In
particular we can calculate the filling fractions $l_1$ and $l_2$ for the two cuts using the general formula of Section 2:
\begin{equation}
 l_j =  \frac{1}{2 \pi } \oint_{\alpha_j} u \: \frac{dt}{t}
\label{eq:filling_fractions_general_formula}
\end{equation}
where $\alpha_j$ is a closed contour encircling the $j$-th cut and no other singularities. In the present case the two cuts correspond to the intervals $[-S/K,0]$ and $[0,S/K]$ in the $u$ plane and the filling fractions are given as:
\begin{eqnarray}
 l = \frac{S}{\pi i} \int_{- \frac{S}{K}}^{0} \frac{du}{\sqrt{u^2 - \frac{S^2}{n^2}}} = \frac{S}{2} \qquad
  \tilde{l} = \frac{S}{\pi i} \int_{0}^{\frac{S}{K}} \frac{du}{\sqrt{u^2 - \frac{S^2}{n^2}}} = \frac{S}{2}
\label{eq:filling_fractions_original_cuts_N-folded_GKP} \nonumber
\end{eqnarray}
and thus turn out to be equal.
\paragraph{}
Finally, to compare the string energy with the dimension of the
corresponding gauge theory operator (\ref{gpred}) we compute the
highest conserved charge 
$q_K$ from (\ref{eq:qktilde_from_spin_chain_N-folded_GKP}):
\begin{equation}
 q_K = (-1)^\frac{K}{2} \left( \frac{2 S}{K} \right)^K
\label{eq:highest_conserved_charge_Nfolded_GKP}
\end{equation}
and then compare \eqref{eq:E_S_omega_to_1_behaviour} with the gauge theory prediction from equation \eqref{gpred}:
\begin{equation}
 \Delta - S = \frac{\sqrt{\lambda}}{2\pi} [\log q_K + C_{\mathrm{string}}(K)]
\label{eq:general_gauge_theory_prediction_for_E-S}
\end{equation}
where we have omitted subleading terms in the limit $\omega\to 1$. We obtain:
\begin{equation}
 \Delta - S = \frac{K \sqrt{\lambda}}{2\pi} \log S + \frac{\sqrt{\lambda}}{2\pi} \left( K \log 2 - K \log K + \log (-1)^\frac{K}{2} 
  + C_{\mathrm{string}}(K) \right)
\label{eq:gauge_theory_prediction_for_E-S_Nfolded_GKP} \nonumber
\end{equation}
which suggests:
\begin{equation}
 C_{\mathrm{string}}(K) = K \left[ \log \left( \frac{8 \pi}{\sqrt{\lambda}} \right) - 1 \right] - \log(-1)^\frac{K}{2}
\label{eq:C(n)_from_Nfolded_GKP}
\end{equation}

We would also like to remark that all calculations concerning the N-folded GKP string reduce to the standard GKP results for $N=1$ (i.e. $K=2$), as listed in \cite{D1,GKP,JJ}.

\subsection{The symmetric spiky string}
\label{sec:Kruc_spiky_string}

\subsubsection{General properties}
\label{sec:general_properties_Kruc}

The Kruczenski spiky string was first discovered \cite{Kruc} as a
solution to the equations of motion generated by the Nambu-Goto
action. More recently, its conformal gauge version was found in 
\cite{JJ}. In appendix \ref{sec:app_gauge_transf_for_K}, we verify 
directly that
the solution presented in \cite{JJ} really is gauge equivalent to the
original spiky string solution of \cite{Kruc}. 
Here, we will only be interested in the conformal
gauge version of the solution. 
The ansatz made in this case is\footnote{As in the previous section 
$\tilde{\sigma}$ denotes the worldsheet coordinate prior to a
  rescaling which normalises its periodicity to $2\pi$.} $t = \tilde{\tau} +
f(\tilde{\sigma})$, $\phi = \phi_0 + \omega \tilde{\tau} +
g(\tilde{\sigma})$, $\rho = \rho(\tilde{\sigma})$, and it leads to the 
following solution to the equations
of 
motion and Virasoro constraints:
\begin{eqnarray}
 \partial_{\tilde{\sigma}} f(\tilde{\sigma}) & = & \frac{\omega \sinh 2 \rho_0}{2 \cosh^2 \rho} \:, \qquad
  \partial_{\tilde{\sigma}} g(\tilde{\sigma}) = \frac{\sinh 2 \rho_0}{2 \sinh^2 \rho} \nonumber \\
 \left[ \partial_{\tilde{\sigma}} \rho(\tilde{\sigma}) \right]^2 & = & \frac{(\cosh^2 \rho - \omega^2 \sinh^2\rho)(\sinh^2 2\rho - \sinh^2 
  2 \rho_0)}{\sinh^2 2\rho}
\label{eq:derivatives_for_JJ_solution}
\end{eqnarray}
Again we impose $\rho_0 \leq \rho \leq \rho_1$, with $\coth\rho_1 = \omega$, so that both factors in the numerator of $[\partial_{\tilde{\sigma}} \rho(\tilde{\sigma})]^2$ are positive. These equations can be integrated to give:
\begin{equation}
 \rho (\tilde{\sigma}) = \frac{1}{2} \cosh^{-1} \: [ \cosh 2\rho_1 \: \mathrm{cn}^2 (v|k) + \cosh 2\rho_0 \: \mathrm{sn}^2 (v|k)]
\label{eq:rho_of_sigmatilde_JJ}
\end{equation}
where
\begin{equation}
 v \equiv \sqrt{\frac{\cosh 2\rho_1 + \cosh 2\rho_0}{\cosh 2 \rho_1 - 1}} \tilde{\sigma} \:, \qquad
  k \equiv \sqrt{\frac{\cosh 2\rho_1 - \cosh 2\rho_0}{\cosh 2 \rho_1 + \cosh 2 \rho_0}}
\label{eq:def_v_and_k_JJ}
\end{equation}
and:
\begin{eqnarray}
 f (\tilde{\sigma}) & = & \frac{\sqrt{2} \omega \sinh 2 \rho_0 \sinh \rho_1}{(\cosh 2 \rho_1 + 1)\sqrt{\cosh 2 \rho_1 + \cosh 2 \rho_0}}
  \Pi \left( \frac{\cosh 2\rho_1 - \cosh 2\rho_0}{\cosh 2\rho_1 + 1}, x, k \right) \nonumber \\
 g (\tilde{\sigma}) & = & \frac{\sqrt{2} \sinh 2 \rho_0 \sinh \rho_1}{(\cosh 2 \rho_1 - 1)\sqrt{\cosh 2 \rho_1 + \cosh 2 \rho_0}}
  \Pi \left( \frac{\cosh 2\rho_1 - \cosh 2\rho_0}{\cosh 2\rho_1 - 1}, x, k \right) \nonumber \\
\label{eq:f_and_g_of_sigmatilde_JJ}
\end{eqnarray}
where $x = \mathrm{am}(v|k)$ ($0 \leq k \leq 1$). For simpler notation, we introduce $w \equiv \cosh 2\rho$, $w_0 \equiv \cosh 2\rho_0$, $w_1 \equiv \cosh 2\rho_1$ and define:
\begin{equation}
 n_\pm \equiv \frac{w_1 - w_0}{w_1 \pm 1} \:, \qquad \mathbb{K} \equiv \mathbb{K}(k) \:, \qquad \mathbb{E} \equiv \mathbb{E}(k)
\label{eq:def_np_nm_JJ}
\end{equation}
which will be used throughout the rest of this section.

In order to understand the shape of this solution, we first need to observe that $\rho(v(\tilde{\sigma}))$ is periodic of period $2\mathbb{K}$, starting off at $\rho(0) = \rho_1$, then decreasing to $\rho(\mathbb{K}) = \rho_0$ at half the period and finally going back to  $\rho(2\mathbb{K}) = \rho_1$. Therefore, for the string to be closed, we impose $v(\tilde{\sigma}) \in [0,2 K \mathbb{K}]$, which corresponds to $\tilde{\sigma} \in [0,L]$, with:
\begin{equation}
 L = 2 K \mathbb{K}\sqrt{\frac{w_1 - 1}{w_1 + w_0}} \equiv 2 K \tilde{L}
\label{eq:range_of_sigmatilde_in_JJ} \nonumber
\end{equation}
$f$ and $g$ are instead pseudo-periodic of pseudo-period $2\tilde{L}$:
\begin{eqnarray}
 f(\tilde{\sigma} + 2 m \tilde{L}) & = & f(\tilde{\sigma}) + \frac{\sqrt{2} \omega \sinh 2 \rho_0 \sinh \rho_1}{(w_1 + 1)\sqrt{w_1 + w_0}} 
  2 m \Pi (n_+,k) \nonumber \\
 g(\tilde{\sigma} + 2 m \tilde{L}) & = & g(\tilde{\sigma}) + \frac{\sqrt{2} \sinh 2 \rho_0 \sinh \rho_1}{(w_1 - 1)\sqrt{w_1 + w_0}} 
  2 m \Pi (n_-,k)
\label{eq:pseudo-periodicities_of_f_and_g_JJ} \nonumber
\end{eqnarray}
due to the pseudo-periodicities of the amplitude function and of the incomplete elliptic integral of the third kind.

Now, in order to have a closed string at constant global time $t$, we need to substitute $\tau = t - f(\tilde{\sigma})$ into the original ansatz for $\phi$, thus finding $\phi(t,\tilde{\sigma}) = \omega t + g(\tilde{\sigma}) - \omega f(\tilde{\sigma})$, and then to impose $\phi(t,L) = \phi(t,0) + 2 n \pi$, for $n\in\mathbb{Z}$. By the pseudo-periodicity, we can easily see that $\phi(t,L) - \phi(t,0) = 2 K \Delta\phi$, where:
\begin{eqnarray}
 \Delta\phi & = & \frac{\sqrt{2}\sinh 2\rho_0 \sinh \rho_1}{\sqrt{w_1+w_0}} \left[ \frac{\Pi(n_-,k)}{w_1 - 1} - \frac{\omega^2\Pi(n_+,k)}
  {w_1 + 1} \right] \nonumber\\
  					& = & \frac{\sinh 2\rho_0}{\sqrt{2} \sinh \rho_1 \sqrt{w_1+w_0}} \left[ \Pi(n_-,k) - \Pi(n_+,k) \right]
\label{eq:deltaphi_for_JJ}
\end{eqnarray}
The closedness constraint then becomes:
\begin{equation}
 \Delta\phi = \frac{n}{K} \pi
\label{eq:closedness_constraint_JJ}
\end{equation}

\begin{figure}% Kruczenski string, 10 spikes
\includegraphics[width=\columnwidth]{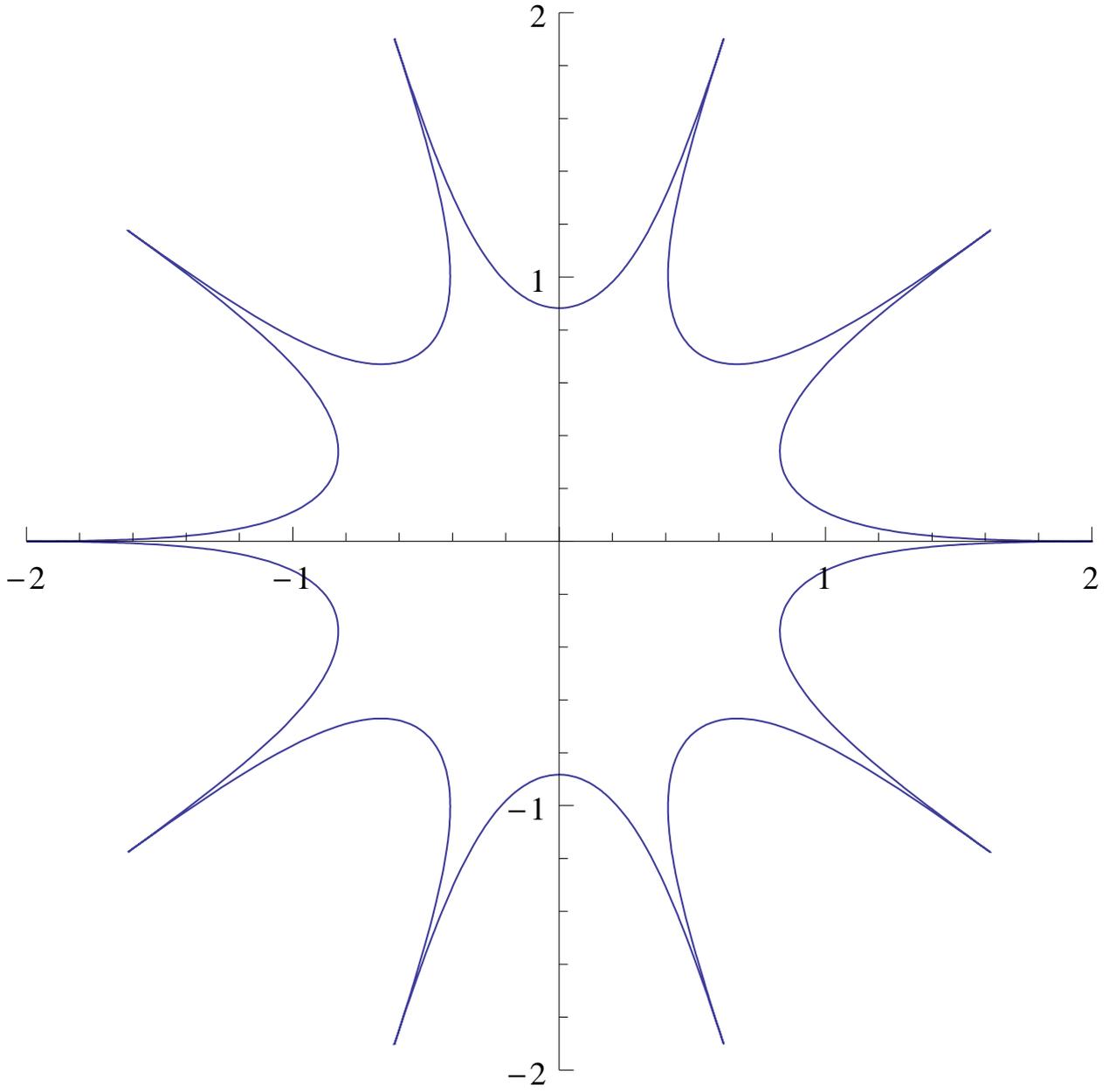}%
\caption{The Kruczenski spiky string in the $(\rho,\phi)$-plane at $t=0$, with $\rho_0 = 0.882663$ and $\rho_1 = 2$.}%
\label{fig:Kruc_10_spikes}%
\end{figure}

The resulting plot is shown in Fig. \ref{fig:Kruc_10_spikes}, and consists of $K$ arcs of equal angular separation $\Delta\theta = 2\Delta\phi$; a cusp is present at the joining point between each pair of consecutive arcs, where $\rho=\rho_1$. As global time varies, the string rigidly rotates.

We can also easily check that the cusp condition is satisfied for $\rho = \rho_1$. Remembering that we're interested in the plot at constant $t$, from \eqref{eq:derivatives_for_JJ_solution} we immediately see that $\partial_{\tilde{\sigma}} \rho = 0$ and $\partial_{\tilde{\sigma}} \phi(t,\tilde{\sigma}) = 0$ for $\rho=\rho_1$. This also implies $\partial_{\tilde{\sigma}} X_\mu = 0$ at constant $t$ for $\rho = \rho_1$, $\mu = 0,\ldots,3$, with $X_\mu$ now representing the embedding coordinates. Therefore, we can deduce that the $n$ cusps are located at:
\begin{equation}
 \tilde{\sigma}_m = 2 m \tilde{L} \:, \qquad m = 0,\ldots,K-1
\label{eq:cusp_positions_JJ_sigmatilde} \nonumber
\end{equation}

As far as the behaviour of the solution in conformal gauge as $\omega\to 1$ is concerned, we find that, analogously to the GKP case, $\rho_1 \to +\infty$, the spikes touch the boundary and, as we'll see shortly, the energy and angular momentum diverge. Note that, when considering this limit, $\rho_0$ is not fixed, since it depends on $\rho_1$ through equations \eqref{eq:deltaphi_for_JJ} and \eqref{eq:closedness_constraint_JJ}. Instead, it changes so that $\Delta\phi$ remains constant.

\begin{figure}%
\includegraphics[width=\columnwidth]{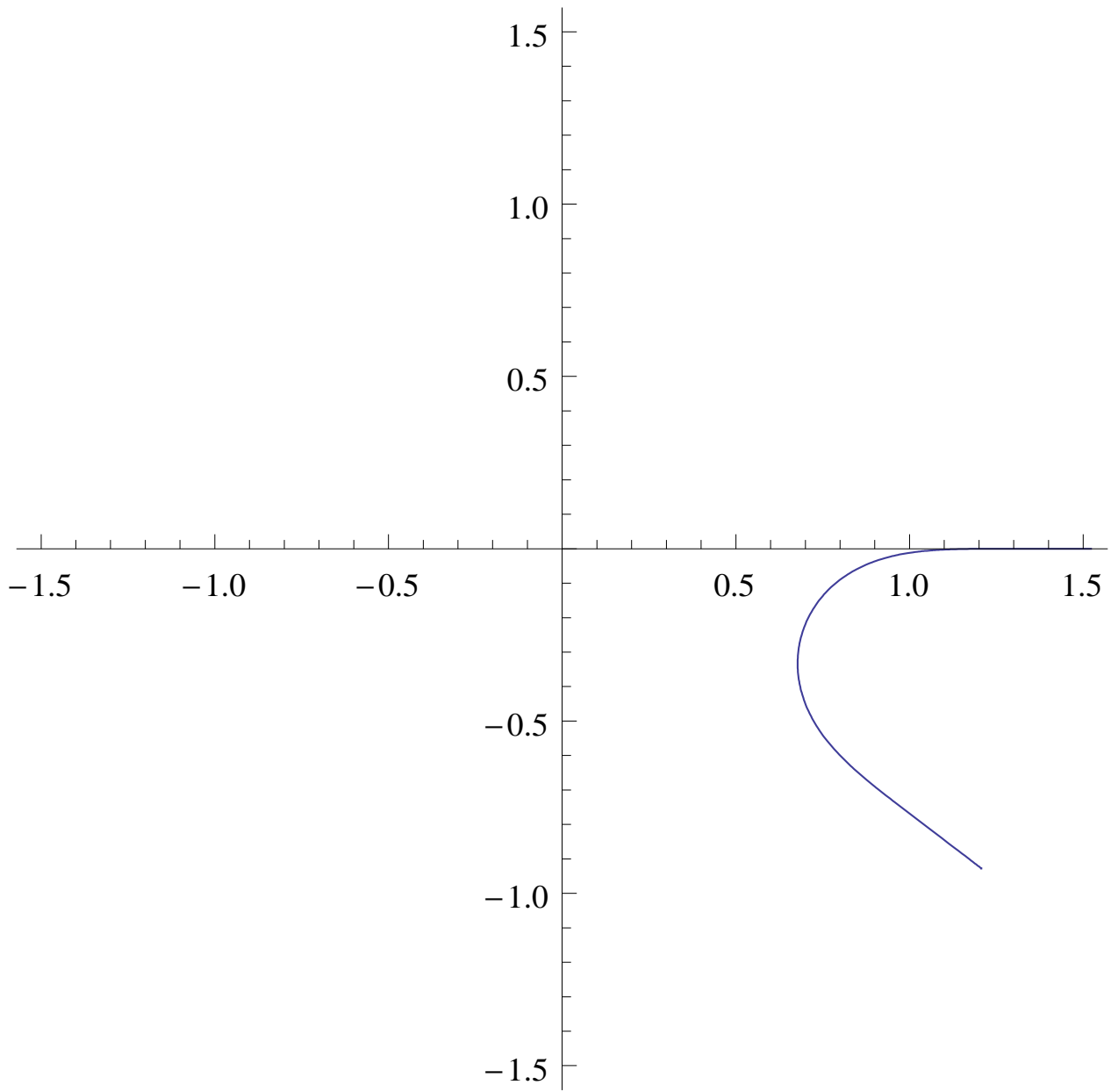}%
\caption{The Kruczenski spiky string in conformal gauge, for $\omega = 1$ and $\rho_0 = 0.9$.}%
\label{fig:JJ_single_arc_rho0=0.9}%
\end{figure}

It is also possible to compute a solution to the equations of motion and Virasoro constraints which holds for $\omega = 1$:
\begin{eqnarray}
 \rho(\tilde{\sigma}) & = & \frac{1}{2} \cosh^{-1} (w_0 \cosh 2\tilde{\sigma}) \nonumber \\
 t(\tilde{\tau},\tilde{\sigma}) & = & \tilde{\tau} + \arctan \left[ \coth 2\rho_0 \: e^{2 \tilde{\sigma}} + \frac{1}{\sinh 2\rho_0} \right]
  \nonumber \\
 \phi(\tilde{\tau},\tilde{\sigma}) & = & \tilde{\tau} + \arctan \left[ \coth 2\rho_0 \: e^{2 \tilde{\sigma}} - \frac{1}{\sinh 2\rho_0} \right]
\label{eq:JJ_single_arc_solution_omega=1}
\end{eqnarray}
This solution is obtained simply by integrating \eqref{eq:derivatives_for_JJ_solution} after substituting in $\omega = 1$. It describes a single arc which has its endpoints on the boundary of $AdS$, reached for $\tilde{\sigma} \to \pm \infty$ (Fig. \ref{fig:JJ_single_arc_rho0=0.9}). What we see is the result of ``blowing up'' one of the interconnecting arcs located between two consecutive cusps in the original $\omega > 1$ solution. In the process, the spikes are ``pushed away'' into the region in which $\tilde{\sigma}$ becomes infinite, ultimately disappearing from the worldsheet. Other than from the plot, we can also see this from the fact that now we have $\left[ \partial_{\tilde{\sigma}} \rho(\tilde{\sigma}) \right]^2 = (\sinh^2 2\rho - \sinh^2 2 \rho_0)/\sinh^2 2\rho$, and thus the first derivative of $\rho(\tilde{\sigma})$ does not vanish any longer at the endpoints $\rho \to +\infty$.

The reason why this solution is particularly helpful is that it allows us to obtain the relationship between the angular separation $\Delta\theta = 2 \Delta\phi$ at constant $t$ of the arcs in the original solution and the parameter $\rho_0$ in the limit $\omega \to 1$. Since \eqref{eq:JJ_single_arc_solution_omega=1} describes one of these arcs at $\omega =1$, all we have to do is to compute $\Delta\theta$ from it:
\begin{equation}
 \Delta\theta = \left( \lim_{\tilde{\sigma} \to +\infty} - \lim_{\tilde{\sigma} \to -\infty} \right) \phi(t,\tilde{\sigma}) = 2 \mathrm{Arctan}
  \frac{1}{\sinh 2\rho_0}
\label{eq:Deltatheta_single_arc_JJ_omega=1} \nonumber
\end{equation}
Therefore, we deduce that the expression defined in \eqref{eq:deltaphi_for_JJ} has the following behaviour:
\begin{equation}
 \Delta\theta \simeq 2 \mathrm{Arctan} \frac{1}{\sinh 2\rho_0} \:, \qquad \textrm{as } \omega \to 1
\label{eq:Deltatheta_single_arc_JJ_omega_to_1}
\end{equation}
Note that $\Delta\theta \in (0,\pi)$, since $0<\rho_0<+\infty$; this is true for any $\omega>1$, due to the fact that, as we remarked earlier, $\Delta\theta$ is always fixed at a constant value by the closedness constraint \eqref{eq:closedness_constraint_JJ}. This implies that $n < K/2$. This result will be helpful later, when computing the monodromy matrix for large $S$, since it shows that $\rho_0$ always approaches a constant non-zero value as $\rho_1$ diverges, and therefore it always behaves as $O(1)$ in the limit $\omega \to 1$.

We can recognise the GKP $N$-folded string solution as a special case
of the spiky string. In particular 
\eqref{eq:Deltatheta_single_arc_JJ_omega_to_1}, shows that, when
$\Delta\theta\to \pi$, we have $\rho_0 \to 0$ and we recover a folded
string solution,
which passes through the origin $\rho=0$. 

\subsubsection{Energy, angular momentum and large angular momentum behaviour}
\label{sec:JJ_E_S_large_S_behaviour}

By using \eqref{eq:rho_of_sigmatilde_JJ}, \eqref{eq:AdS3_energy} and \eqref{eq:AdS3_spin}, we can easily compute the energy and the angular momentum of this solution:
\begin{eqnarray}
 \Delta & = & K \frac{\sqrt{\lambda}}{2\pi} \int_0^{2\tilde{L}} d\tilde{\sigma} \cosh^2 \rho (\tilde{\sigma}) \nonumber \\
   & = & K \frac{\sqrt{\lambda}}{\pi} \sqrt{\frac{w_1-1}{w_1+w_0}} \left[ \frac{1}{2} (w_1+w_0) \mathbb{E} - \sinh^2 \rho_0 \mathbb{K} \right] 
   \label{eq:E_JJ}\nonumber\\
 S & = & K \frac{\omega\sqrt{\lambda}}{2\pi} \int_0^{2\tilde{L}} d\tilde{\sigma} \sinh^2 \rho (\tilde{\sigma}) \nonumber \\
   & = & K \frac{\omega\sqrt{\lambda}}{\pi} \sqrt{\frac{w_1-1}{w_1+w_0}} \left[ \frac{1}{2} (w_1+w_0) \mathbb{E} - \cosh^2 \rho_0 \mathbb{K}
    \right] 
   \label{eq:S_JJ}\nonumber
\end{eqnarray}
both of which are just $K$ times the contribution of a single arc, due to the periodicity of $\rho(\tilde{\sigma})$.

Again, we consider the limit as $\omega \to 1$, with $\omega = 1+\eta$, $\eta \gtrsim 0$, and expand these two quantities:
\begin{eqnarray}
 \Delta & = & K \frac{\sqrt{\lambda}}{2 \pi \eta} - K \frac{\sqrt{\lambda}}{32\pi} (8+3w_0) \log \eta \nonumber \\
   &   & + K \frac{\sqrt{\lambda}}{64\pi} \left[ -13w_0 + (32+12w_0) \log \frac{2 \sqrt{2}}{\sqrt{w_0}} \right] + O(\eta\log\eta)
           \label{eq:behaviour_of_E_JJ_as_omega_goes_to_1} \\
 S & = & K \frac{\sqrt{\lambda}}{2 \pi \eta} + K \frac{\sqrt{\lambda}}{32\pi} (8-3w_0) \log \eta \nonumber \\
   &   & + K \frac{\sqrt{\lambda}}{64\pi} \left[32 -13w_0 + (-32+12w_0) \log \frac{2 \sqrt{2}}{\sqrt{w_0}} \right] + O(\eta\log\eta)
           \label{eq:behaviour_of_S_JJ_as_omega_goes_to_1}
\end{eqnarray}
and we still find that each spike contributes $\sqrt{\lambda}/(2\pi\eta)$ to the leading behaviour in both cases. We also compute the $O(1)$ correction to the anomalous dimension:
\begin{equation}
 \Delta - S = \frac{K\sqrt{\lambda}}{2\pi} \log \left( \frac{2\pi S}{K\sqrt{\lambda}} \right) + \frac{K\sqrt{\lambda}}{2\pi} \left[ 3\log 2 - 1 +
  \log \left( \sin \frac{\Delta\theta}{2} \right) \right] + O(\eta\log\eta)
\label{eq:E-S_omega_to_1_behaviour}
\end{equation}
where we have used the relation:
\begin{equation}
 \frac{1}{w_0} \simeq \sin \frac{\Delta\theta}{2} \qquad \textrm{as } \omega \to 1
\label{eq:relation_between_u0_and_sin_Deltatheta/2}
\end{equation}
which is easily obtained from
\eqref{eq:Deltatheta_single_arc_JJ_omega_to_1}. 
Again we find the usual logarithmic growth $\Delta-S\sim K\log(S)$
which is characteristic of the gauge theory anomalous dimensions for
operators of twist $K$. 

\subsubsection{Spectral curve for large S}
\label{sec:spectral_curve_JJ}

We'll now repeat the calculation carried out in section \ref{sec:spectral_curve_Nfolded_GKP} for the Kruczenski spiky string in conformal gauge.

As before, we start by computing the components of the right current, by substituting \eqref{eq:rho_of_sigmatilde_JJ} into \eqref{eq:components_of_j_tau_in_terms_of_t_phi_rho}, and then re-expressing them in terms of the rescaled worldsheet coordinates $(\tau,\sigma)$, defined so that $\sigma \in [0,2\pi]$:
\begin{equation}
 (\tau,\sigma) = \frac{2\pi}{L} (\tilde{\tau},\tilde{\sigma}) = \frac{\pi}{K \mathbb{K}}
  \sqrt{\frac{w_1+w_0}{w_1-1}}(\tilde{\tau},\tilde{\sigma})
\label{eq:def_rescaled_coords_JJ} \nonumber
\end{equation}
The charge density is then given by:
\begin{eqnarray}
 j^0_\tau(\tau,\sigma) & = & \frac{K \mathbb{K}}{\pi}\sqrt{\frac{w_1-1}{w_1+w_0}} \nonumber \\
  & & \times \{ (\omega+1) [w_1 \mathrm{cn}^2 (v|k) + w_0 \mathrm{sn}^2 (v|k)] + 1-\omega \} \nonumber \\
 j^1_\tau(\tau,\sigma) + i j^2_\tau(\tau,\sigma) & = & i \frac{K \mathbb{K}}{\pi} \sqrt{\frac{w_1-1}{w_1+w_0}} (\omega+1) e^{i(\phi-t)}  
  \nonumber \\
  & & \times \sqrt{[w_1 \mathrm{cn}^2 (v|k) + w_0 \mathrm{sn}^2 (v|k)]^2 - 1}
\label{eq:eq:j_tau_in_terms_of_tau_sigma_JJ}
\end{eqnarray}
where now:
\begin{equation}
 v \equiv \sqrt{\frac{w_1 + w_0}{w_1 - 1}} \tilde{\sigma} = \frac{K \mathbb{K}}{\pi} \sigma
\label{eq:v_in_terms_of_sigma_JJ} \nonumber
\end{equation}
In terms of the new coordinates, the cusps are located at:
\begin{equation}
 \sigma_m = 2m \frac{\pi}{K} \:, \qquad m=0,\ldots,K-1
\label{eq:cusp_positions_JJ_sigma}
\end{equation}
As in the GKP case, the leading order of the charge density will be dominated by the contributions coming from the cusps, which we compute individually by setting $\sigma = \sigma_m + \hat{\sigma}$, with $|\hat{\sigma}| < \pi/K$, and then expanding \eqref{eq:eq:j_tau_in_terms_of_tau_sigma_JJ} as $\omega \to 1$, obtaining:
\begin{equation}
 j_\tau^0 (\tau,\sigma) \simeq  \frac{2 K \mathbb{K}}{\pi} \frac{1}{\eta \cosh^2 \left( \frac{K \mathbb{K}}{\pi} \hat{\sigma} \right)}
\label{eq:leading_behaviour_of_jtau0_JJ} \nonumber
\end{equation}
which is the same we found in the N-folded GKP case (c.f. equation \eqref{eq:leading_behaviour_for_cpts_of_j_tau_GKP_Nfolded}), except for the fact that $\mathbb{K} \equiv \mathbb{K}(1/\omega)$ for GKP, whereas now we have $\mathbb{K} \equiv \mathbb{K}(k)$ (the leading behaviour $\mathbb{K} \simeq -(1/2) \log \eta$ is however identical).

For the remaining two components of the charge density we find:
\begin{equation}
 j^1_\tau(\tau,\sigma) + i j^2_\tau(\tau,\sigma) \simeq i \frac{2 K \mathbb{K}}{\pi} \frac{1}{\eta \cosh^2 \left( \frac{K \mathbb{K}}{\pi}
  \hat{\sigma} \right)} e^{i \left( \phi_0 + m \frac{2n\pi}{K} \right)}
\label{eq:leading_behaviour_of_jtau1+ijtau2_JJ} \nonumber
\end{equation}

We can now use definition \eqref{eq:def_normalised_charge_density}, together with the leading behaviour of $S$ as $\omega \to 1$, which can be extracted from \eqref{eq:behaviour_of_S_JJ_as_omega_goes_to_1}, and with identity \eqref{eq:identity_for_delta_function}, to calculate the contribution to the normalised charge density from the $m$-th cusp. Then, we only have to sum over all values of $m$ to obtain:
\begin{eqnarray}
 \mu^0 (\tau,\sigma) & = & \frac{1}{K} \sum_{m=0}^{K-1} \delta (\sigma - \sigma_m) \nonumber\\
 \mu^1 (\tau,\sigma) + i \mu^2 (\tau,\sigma) & = & i \frac{1}{K} \sum_{m=0}^{K-1} e^{i \left( \phi_0 + m \frac{2n\pi}{K} \right)} 
  \delta (\sigma - \sigma_m)
\label{eq:cpts_of_mu_JJ} \nonumber
\end{eqnarray}
From these equations, we can calculate the spin vector at each spike according to \eqref{eq:def_spin_vector_at_each_cusp}:
\begin{equation}
 \vec{L}_m = \frac{S}{K} \begin{pmatrix}
	                  1 \\
	                  -\sin \left(\phi_0 + m \frac{2 n \pi}{K} \right) \\
	                  \cos \left(\phi_0 + m \frac{2 n \pi}{K} \right)
                   \end{pmatrix}
\label{eq:spin_vector_at_mth_cusp_JJ} \nonumber
\end{equation}
As in the GKP case, these vectors satisfy the properties listed in \eqref{eq:properties_of_spin_vectors}, so that the Kruczenski solution is also a highest weight state.

Then, we can evaluate the monodromy matrix as we did in the folded
string case, leading to  
 \eqref{eq:monodromy_matrix_as_omega_to1_utilde}, where in this case the matrix $\mathbb{L}_m(u)$ reads:
\begin{equation}
 \mathbb{L}_m (u) = \begin{pmatrix}
	                           u + \frac{i S}{K}   & \frac{i S}{K} e^{i\left(\phi_0 + m \frac{2 n \pi}{K} \right)} \\
	                           -\frac{i S}{K} e^{-i\left(\phi_0 + m \frac{2 n \pi}{K} \right)} & u - \frac{i S}{K}
                            \end{pmatrix} % \equiv \begin{pmatrix}
	                                        %       a & b_m \\
	                                        %       \bar{b}_m & \bar{a}
                                          %        \end{pmatrix}
\label{eq:L_m(utilde)_for_JJ_1}
\end{equation}
We notice that this time no simplification occurs, i.e. the product of two consecutive matrices $\mathbb{L}_m (u)$ and $\mathbb{L}_{m+1} (u)$ still depends on $m$, and therefore we can't proceed as we did earlier. Instead, we introduce the following sequence of matrices:
\begin{equation}
 S_m = \begin{pmatrix}
	      c e^{i m \frac{n \pi}{K}} & d e^{i m \frac{n \pi}{K}} \\
	      \bar{d} e^{-i m \frac{n \pi}{K}} & \bar{c} e^{-i m \frac{n \pi}{K}}
       \end{pmatrix} \:, \qquad \textrm{with } |c|^2 - |d|^2 = 1
\label{eq:acceptable_choice_for_S_m_JJ} \nonumber
\end{equation}
for $m=0,\ldots,K$ (where $c$ and $d$ are arbitrary, apart from the constraint on their absolute values), and notice that it makes the product $S_m^{-1} \mathbb{L}_m (u) S_{m+1} \equiv \mathbb{M} (u)$ independent of $m$. We also observe that:
\begin{equation}
 S_n = \begin{pmatrix}
	      c e^{i n \pi} & d e^{i n \pi} \\
	      \bar{d} e^{-i n \pi} & \bar{c} e^{-i n \pi}
       \end{pmatrix} = (-1)^{n} \begin{pmatrix}
	                                       c & d \\
	                                       \bar{d} & \bar{c}
                                        \end{pmatrix} = (-1)^{n} S_0
\label{eq:relation_between_S_n_and_S_0_JJ}
\end{equation}
We can now compute the trace of the monodromy matrix, by inserting copies of the identity matrix, in the form of the products $S_m S_m^{-1}$, between consecutive matrices $\mathbb{L}_m (u)$:
\begin{eqnarray}
 \mathrm{tr} \, \Omega[x] & = & \frac{1}{u^K} \mathrm{tr} \prod_{m=0}^{K-1} \mathbb{L}_m(u) \nonumber\\
  & = & \frac{1}{u^K} \mathrm{tr} \left[ S_0 S_0^{-1} \mathbb{L}_0 (u) S_1 S_1^{-1} \mathbb{L}_1(u) S_2 S_2^{-1}
   \ldots S_{K-1} S_{K-1}^{-1} \mathbb{L}_{K-1}(u) \right] \nonumber \\
  & = &  \frac{(-1)^{n}}{u^K} \mathrm{tr} \left[ S_0^{-1} \mathbb{L}_0 (u) S_1 S_1^{-1} \mathbb{L}_1(u) S_2
   S_2^{-1} \ldots S_{K-1} S_{K-1}^{-1} \mathbb{L}_{K-1}(u) S_K \right] \nonumber \\
  & = & (-1)^{n} \frac{1}{u^K} \mathrm{tr} \left[ \mathbb{M} (u)^K \right]
\label{eq:trace_of_monodromy_matrix_in_terms_of_M_utilde_JJ} \nonumber
\end{eqnarray}
where in obtaining the third line we have used \eqref{eq:relation_between_S_n_and_S_0_JJ} and the cyclicity property of the trace.

The rest of the calculation proceeds as in the GKP case. We first determine the eigenvalues of $\mathbb{M} (u)$:
\begin{equation}
 \kappa_\pm = u \cos \frac{n\pi}{K} - \frac{S}{K} \sin \frac{n\pi}{K} \pm \sqrt{- \frac{2 S}{K} u 
  \sin \frac{n\pi}{K} \cos \frac{n\pi}{K} + \left( \frac{S^2}{K^2} - u^2 \right) \sin^2 \frac{n\pi}{K}}
\label{eq:evalues_of_M(utilde)_JJ} \nonumber
\end{equation}
and then deduce:
\begin{eqnarray}
 \mathrm{tr}\: \Omega[x] & = & (-1)^{n} \frac{1}{u^K} 
  ( \kappa_+^K + \kappa_-^K ) \nonumber \\
  & = & (-1)^{n} 2 T_K \left( \cos \frac{n\pi}{K}  - \frac{ S \sin \frac{n\pi}{K}}{K u} \right) \nonumber \\
  & = & 2 \cos \left[ n \pi + K \cos^{-1} \left( \cos \frac{n\pi}{K}  - \frac{S \sin \frac{n\pi}{K}}{K u} \right)
   \right]
\label{eq:tr_Omega_JJ} \nonumber
\end{eqnarray}
where again we have used \eqref{eq:def_Tk(y)_Chebyshev}. Hence, we obtain the following expression for the quasi-momentum:
\begin{equation}
 p(u) = n \pi + K \cos^{-1} \left( \cos \frac{n\pi}{K}  - \frac{S \sin \frac{n\pi}{K}}{K u} \right)
\label{eq:quasi-momentum_K}
\end{equation}
which then yields the expression for the spectral curve for the Kruczenski solution:
\begin{eqnarray}
 \Sigma_{\Omega}\,\,: \qquad{} \qquad{} t\,\,+\,\,\frac{1}{t} & = & 
2 \cos \left[
  n \pi + K \cos^{-1} \left( \cos \frac{n\pi}{K}  - \frac{S \sin \frac{n\pi}{K}}{K u} \right) \right] 
\label{eq:string_spectral_curve_K}
\end{eqnarray}
We see that this again corresponds to a point in the moduli space of the curve $\Gamma_K$ appearing in equation \eqref{eq:general_curve_from_spin_chain} where the conserved charges take the following values:
\begin{equation}
 q_k = \left( - \frac{2 S}{K} \right)^k \sum_{1\leq j_1 < j_2 < \ldots < j_k \leq K} \prod_{l=1}^k \sin \left[ \frac{n\pi}{K}
  (j_{l+1} - j_l) \right] \:, \quad k=2, \ldots, K
\label{eq:qktilde_from_spin_chain_K}
\end{equation}
where $j_{k+1} \equiv j_1$ (the normalisation $q_2 = -S^2$ is checked in appendix \ref{sec:app_computing_qtilde_2_for_N-folded_GKP_and_K}).
\paragraph{}
We can now compute the discriminant $D = 4 \sin^2 p(u)$:
\begin{eqnarray}
 D (u) & = & 4 \left[1 - T_K^2 \left( \cos \frac{n\pi}{K} - \frac{S \sin
  \frac{n\pi}{K}}{K u} \right) \right] = \nonumber \\
 & = & - 4 \sin \frac{n\pi}{K} \left(- \sin \frac{n\pi}{K} - \frac{2 S}{K u} \cos \frac{n\pi}{K} +
  \frac{S^2}{K^2 u^2} \sin \frac{n\pi}{K} \right) \nonumber \\
 & & \times U_{K-1}^2 \left( \cos \frac{n\pi}{K} - \frac{S \sin
  \frac{n\pi}{K}}{K u} \right) %\nonumber \\
\label{eq:rewriting_D(utilde)_JJ} \nonumber
\end{eqnarray}
which we use to determine the pattern of branch points for the spectral curve \eqref{eq:string_spectral_curve_K}:
\begin{equation}
 \begin{array}{lc}
  u^{-1}  =  \frac{K}{S \sin \frac{n\pi}{K}} \left( \cos \frac{n\pi}{K} \pm 1 \right)  & \qquad \textrm{simple} \\
  u^{-1}  =  \frac{K}{S \sin \frac{n\pi}{K}} \left( \cos
    \frac{n\pi}{K} - \cos \frac{j\pi}{K} \right) \:, \qquad \textrm{for } j=1,\ldots,K-1 & \qquad \textrm{double}
 \end{array}
\label{eq:pattern_of_zeros_of_Q_n_for_JJ}
\end{equation}
As a side remark, we observe that, for the special case of even $K$, these zeros can be rewritten in a form which is very similar to that of the GKP zeros \eqref{eq:pattern_of_zeros_of_Q_n_for_Nfolded_GKP}:
\begin{equation}
 \begin{array}{lc}
  u^{-1}  =  \frac{K}{S \sin \frac{n\pi}{K}} \left( \cos \frac{n\pi}{K} \pm 1 \right)  & \qquad \textrm{simple} \\
  u^{-1}  =  \frac{K}{S \sin \frac{n\pi}{K}} \cos \frac{n\pi}{K} & \qquad \textrm{double} \\
  u^{-1}  =  \frac{K}{S \sin \frac{n\pi}{K}} \left( \cos \frac{n\pi}{K} \pm \sin \frac{l \pi}{K} \right) \:, \qquad
   \textrm{for } l=1,\ldots,\frac{K}{2} - 1 & \qquad \textrm{double}
 \end{array}
\label{eq:pattern_of_zeros_of_Q_n_for_JJ_n_even} \nonumber
\end{equation}
these are exactly the same values appearing in \eqref{eq:pattern_of_zeros_of_Q_n_for_Nfolded_GKP}, shifted by $\cos \frac{n\pi}{K}$ and rescaled by $K/(S \sin \frac{n\pi}{K})$.

As before, we have $K-1$ double zeros and thus the spectral curve degenerates to genus zero. Accordingly, the quasi-momentum \eqref{eq:quasi-momentum_K} is again an analytic function on the complex plane, with a logarithmic branch point at $u = 0$ and two square root branch points at:
\begin{equation}
 u = \frac{S \sin \frac{n\pi}{K}}{K \left( \cos \frac{n\pi}{K} \pm 1 \right)} \equiv u_\pm
\label{eq:sqrt_branch_points_of_p(x)_K} \nonumber
\end{equation}
which always satisfy $u_- < 0 < u_+$. In order to make it single-valued, we introduce the usual branch cut connecting $u_\pm$ and the origin along the real axis. The differential $dp(u)$ is given by:
\begin{equation}
 dp(u) = - \frac{K \sin \frac{n\pi}{K} du}{u \sqrt{\left(\frac{K u}{S} \sin \frac{n\pi}{K} + 
  \cos \frac{n\pi}{K} \right)^2 - 1}}
\label{eq:dp(x)_K} \nonumber
\end{equation}
and displays a simple pole at $u=0$, two square root branch points at $u=u_\pm$ and the other two corresponding square root branch points at infinity. As in the GKP case, the same cut we introduced for $p(u)$ also makes $dp(u)$ single-valued.
\paragraph{}
On the gauge theory side, this spectral curve can again originate from a two-cut solution with cuts colliding at the origin $u=0$. These cuts correspond to the intervals $[u_-,0]$ and $[0,u_+]$, and, according to equation \eqref{eq:filling_fractions_general_formula}, their filling fractions are:
\begin{eqnarray}
 l & = & \frac{1}{\pi i} \int_{u_-}^{0} \frac{K \sin \frac{n\pi}{K} du}{\sqrt{\left( \frac{K u}{S}
  \sin \frac{n\pi}{K} + \cos \frac{n\pi}{K} \right)^2 - 1}} = S \left( 1 - \frac{n}{K} \right) \nonumber \\
 \tilde{l} & = & \frac{1}{\pi i} \int_{0}^{u_+} \frac{K \sin \frac{n\pi}{K} du}{\sqrt{\left( \frac{K u}{S}
  \sin \frac{n\pi}{K} + \cos \frac{n\pi}{K} \right)^2 - 1}} = S \frac{n}{K}
\label{eq:filling_fractions_original_cuts_K} \nonumber
\end{eqnarray}
We observe that, differently from the GKP case, there is an asymmetry in the filling fractions, which is clearly due to the fact that the two square root branch points in $p(u)$ are no longer symmetric with respect to the origin.

Finally, as we did in the previous case, we compare the string energy with the dimension of the corresponding gauge theory operator \eqref{gpred}, by computing the highest conserved charge:
\begin{equation}
 q_K = (-1)^{K+ n} \left( \frac{2 S}{K} \right)^K \left( \sin \frac{n\pi}{K} \right)^K
\label{eq:highest_conserved_charge_JJ}
\end{equation}
Thus, the gauge theory prediction \eqref{eq:general_gauge_theory_prediction_for_E-S} for $\Delta-S$ gives the following expression:
\begin{eqnarray}
 \Delta - S & = & \frac{K\sqrt{\lambda}}{2\pi} \log S + \frac{\sqrt{\lambda}}{2\pi} \left[ K \log 2 
  - K \log K + \log (-1)^{K+n} + K \log \left( \sin \frac{n\pi}{K} \right) \right. \nonumber\\
 & & \left. \phantom{\left( \sin \frac{n\pi}{K} \right)} + C_{\mathrm{string}}(K) \right]
\label{eq:gauge_theory_prediction_for_E-S_JJ} \nonumber
\end{eqnarray}
where again we have omitted terms which are subleading as $\omega \to 1$. Comparison with \eqref{eq:E-S_omega_to_1_behaviour} (we recall that $\Delta\theta = 2 \Delta\phi = 2 (n/K) \pi$) yields:
\begin{equation}
 C_{\mathrm{string}}(K) = K \left[ \log \left( \frac{8 \pi}{\sqrt{\lambda}} \right) - 1 \right] - \log (-1)^{K + n}
\label{eq:C(n)_from_JJ}
\end{equation}

Furthermore, we would like to observe that it is possible to obtain all results for the GKP $N$-folded string from the Kruczenski string in conformal gauge, if we assume, of course, that the two solutions have the same number of cusps. This is done by interpreting the GKP configuration as a set of $K=2N$ spikes with angular separation between consecutive cusps equal to $\pi$, or, in other words, a set of 2 spikes for each turn around the origin in $AdS$ space, which can then be described by a Kruczenski-type solution with $K$ even and $n = K/2$ (which implies $\Delta\theta = \pi$).

First of all, we can substitute this into \eqref{eq:E-S_omega_to_1_behaviour} to recover \eqref{eq:E_S_omega_to_1_behaviour}. \eqref{eq:behaviour_of_E_JJ_as_omega_goes_to_1} and \eqref{eq:behaviour_of_S_JJ_as_omega_goes_to_1} are also seen to reduce to \eqref{eq:E_behaviour_as_omega_goes_to_1_Nfolded_GKP} and \eqref{eq:S_behaviour_as_omega_goes_to_1_Nfolded_GKP} respectively, by noticing that $w_0 = 1$ in the limit $\omega\to 1$ for $\Delta\theta = \pi$, which is easily deduced from \eqref{eq:relation_between_u0_and_sin_Deltatheta/2}. Moreover, all the conserved charges listed in equations \eqref{eq:qktilde_from_spin_chain_K} and \eqref{eq:qktilde_from_spin_chain_N-folded_GKP} also match. Lastly, \eqref{eq:C(n)_from_JJ} agrees with \eqref{eq:C(n)_from_Nfolded_GKP}.

\section{The general patched solution}
\label{sec:patching_K}

\subsection{General properties}
\label{sec:general_properties_patched_K}

In this section, we're going to discuss a generalised version of
Kruczenski's solution in conformal gauge, which allows for arcs with
arbitrary individual angular separations $\Delta\theta_j$,
$j=1,\ldots,K$, which however are still subject to the constraint
$\Delta\theta_j \in (0,\pi)$ (as seen at the end of section
\ref{sec:general_properties_Kruc}; this is an intrinsic property of
the solution found by Jevicki and Jin). The idea is to use different
versions of \eqref{eq:rho_of_sigmatilde_JJ} and
\eqref{eq:f_and_g_of_sigmatilde_JJ} to describe each single arc, and
then to patch all the arcs together by gluing them at the
endpoints. In this way, we are going to construct an approximate
solution, which becomes exact in the 
large angular momentum limit $\omega \to 1$.

We start by considering equation \eqref{eq:deltaphi_for_JJ}, which
determines the angular separation between two consecutive
cusps, as a function of the two parameters $\rho_0$ and $\rho_1$. We
keep $\rho_1$ fixed, and define $K$ parameters $\rho_0^{(j)}$,
$j=1,\ldots,K$ 
by imposing the following constraints:
\begin{equation}
 \Delta\theta_j = \frac{\sqrt{2}\sinh 2\rho_0^{(j)}}{\sinh \rho_1 \sqrt{w_1+w_0^{(j)}}} \left[ \Pi(n_-^{(j)},k^{(j)}) - \Pi(n_+^{(j)},k^{(j)})
  \right] \:, \qquad \textrm{for } j=1,\ldots,K
\label{eq:def_rho_0^j_patched_K}
\end{equation}
where $n_\pm^{(j)}$ and $k^{(j)}$ (and also $v^{(j)}$, which we'll use later) are defined in \eqref{eq:def_v_and_k_JJ} and \eqref{eq:def_np_nm_JJ}, with $\rho_0$ replaced by $\rho_0^{(j)}$. Each pair $(\rho_0^{(j)},\rho_1)$ defines a different version of the solution given in \eqref{eq:rho_of_sigmatilde_JJ} and \eqref{eq:f_and_g_of_sigmatilde_JJ}, with different fundamental half-period $\mathbb{K}_j \equiv \mathbb{K}(k^{(j)})$ (we also define $\mathbb{E}_j \equiv \mathbb{E}(k^{(j)})$) and angular separation $\Delta\theta_j$, but with the same radial position of the spikes $\rho=\rho_1$. We also define:
\begin{equation}
 \tilde{L}_j = \mathbb{K}_j \sqrt{\frac{w_1 - 1}{w_1+w_0^{(j)}}} \:, \qquad \tilde{\sigma}_j = 2 \sum_{k=1}^j \tilde{L}_k 
  \:, \qquad L = 2 \sum_{j=1}^K \tilde{L}_j
\label{eq:def_Ltilde_j_sigmatilde_j_and_L_patched_K}
\end{equation}
In order to glue these different solutions together, we let
$\tilde{\sigma}$ run in the interval $[0,L]$: for $0 =
\tilde{\sigma}_0 \leq \tilde{\sigma} \leq \tilde{\sigma}_1$ we want
the patched solution to describe the first period of the
$(\rho_0^{(1)},\rho_1)$ spiky string, for $\tilde{\sigma}_1 \leq
\tilde{\sigma} \leq \tilde{\sigma}_2$ we want it to describe the first
period of the $(\rho_0^{(2)},\rho_1)$ spiky string, and so on until we
see the first period of the $(\rho_0^{(K)},\rho_1)$ spiky string for
$\tilde{\sigma}_{K-1} \leq \tilde{\sigma} \leq L$. This is achieved by
the 
following definition:
\begin{eqnarray}
 \rho (\tilde{\sigma}) & = & \rho ( \tilde{\sigma}- \tilde{\sigma}_{j-1}, \rho_0^{(j)} ) \nonumber\\
 f (\tilde{\sigma}) & = & f ( \tilde{\sigma}- \tilde{\sigma}_{j-1}, \rho_0^{(j)} ) + \sum_{k=1}^{j-1} f(2\tilde{L}_k,\rho_0^{(k)}) \nonumber\\
 g (\tilde{\sigma}) & = & g ( \tilde{\sigma}- \tilde{\sigma}_{j-1}, \rho_0^{(j)} ) + \sum_{k=1}^{j-1} g(2\tilde{L}_k,\rho_0^{(k)}) \:,
  \nonumber\\
 & & \qquad\qquad\qquad\qquad\qquad \textrm{for } \tilde{\sigma}_{j-1} \leq \tilde{\sigma} \leq \tilde{\sigma}_j
\label{eq:def_patched_K_solution}
\end{eqnarray}
where $\rho (\tilde{\sigma},\rho_0)$, $f (\tilde{\sigma},\rho_0)$ and $g (\tilde{\sigma},\rho_0)$ are given by equations \eqref{eq:rho_of_sigmatilde_JJ} and \eqref{eq:f_and_g_of_sigmatilde_JJ}. In obtaining this, we have used our freedom to shift $\tilde{\sigma}$, $f$ and $g$ by a constant (these are all symmetries of equations \eqref{eq:derivatives_for_JJ_solution}).

We note that, although \eqref{eq:def_patched_K_solution} clearly satisfies the equations of motion and Virasoro constraints in each interval $\tilde{\sigma}_{j-1} < \tilde{\sigma} < \tilde{\sigma}_j$, it is not smooth at the junction points. In particular, $\rho (\tilde{\sigma})$ is $C^1$, whereas $f (\tilde{\sigma})$ and $g (\tilde{\sigma})$ are only $C$. In the case of $\rho$, we can see this by considering equation \eqref{eq:derivatives_for_JJ_solution}:
\begin{eqnarray}
 && \partial_{\tilde{\sigma}} \rho (\tilde{\sigma}) = \pm \sqrt{h(\rho)} \:, \qquad h(\rho) = h_1 (\rho) h_2(\rho) \nonumber \\
 && h_1(\rho) = \cosh^2 \rho - \omega^2 \sinh^2 \rho \:, \qquad h_2 (\rho) = 1 - \frac{\sinh^2 2 \rho_0^{(j)}}{\sinh^2 2 \rho} \nonumber \\
 && \qquad \qquad \qquad \qquad \qquad \qquad \textrm{for } \tilde{\sigma}_{j-1} \leq \tilde{\sigma} \leq \tilde{\sigma}_j 
\label{eq:def_h_h1_h2_studying_derivs_of_rho_patched_K}
\end{eqnarray}
where the sign is plus or minus depending on which half of the $j$-th arc we are considering ($\rho$ is an increasing function of $\tilde{\sigma}$ along one half of every arc and it is instead decreasing on the other half). Clearly, at the junction points $\rho = \rho_1$ we have $\partial_{\tilde{\sigma}} \rho (\tilde{\sigma}) = 0$, independently of $j$, due to the fact that $h_1(\rho_1) = 0$. However, when we turn our attention to the second derivative of $\rho$, we obtain:
\begin{equation}
 \partial_{\tilde{\sigma}}^2 \rho (\tilde{\sigma}) = \frac{h'(\rho)}{2} \:, \qquad \textrm{where } ' \equiv \partial_\rho
\label{eq:rho''_patched_K}
\end{equation}
and it is very easy to see that $h'(\rho)$ contains a term which does not vanish at $\rho = \rho_1$ and which depends on $\rho_0^{(j)}$:
\begin{equation}
 h'(\rho_1) = h_1' (\rho_1) h_2(\rho_1) = (1- \omega^2) \sinh 2 \rho_1 \left( 1 - \frac{\sinh^2 2 \rho_0^{(j)}}{\sinh^2 2 \rho_1} \right)
\label{eq:discontinuity_term_in_rho''_patched_K} \nonumber
\end{equation}
This term generates a discontinuity in $\partial_{\tilde{\sigma}}^2 \rho (\tilde{\sigma})$ at the junction points, since the value of $\rho_0$ jumps from $\rho_0^{(j)}$ to $\rho_0^{(j+1)}$ there.

Similarly:
\begin{equation}
 \partial_{\tilde{\sigma}} g (\tilde{\sigma}) = \frac{\sinh 2 \rho_0^{(j)}}{2} l(\rho) \:, \qquad l(\rho) = \frac{1}{\sinh^2 \rho}  
\label{eq:def_l_in_g'_patched_K} \nonumber
\end{equation}
explicitly depends on $\rho_0^{(j)}$ at $\rho = \rho_1$ and thus is discontinuous at the junction points. The same clearly applies to $\partial_{\tilde{\sigma}} f (\tilde{\sigma})$.

Therefore, the patched version of Kruczsenki's solution is not a proper closed string solution for fixed $\omega>1$. However, the situation changes as $\omega \to 1$. In fact, in this limit, the function $\rho(\tilde{\sigma})$ from equation \eqref{eq:rho_of_sigmatilde_JJ} displays the following leading behaviour near the cusp located at $\tilde{\sigma}=0$:
\begin{equation}
 \rho(\tilde{\sigma}) = - \frac{1}{2} \log \eta + \frac{1}{2} \log ( 2 \mathrm{sech}^2 \tilde{\sigma} ) + O(\eta)
\label{eq:leading_universal_behaviour_of_rho_near_cusps_patched_K} \nonumber
\end{equation}
where, in our usual notation, $\omega = 1 + \eta$. Clearly, the situation is identical near any other cusp, due to the periodicity of $\rho(\tilde{\sigma})$. Thus, $\rho (\tilde{\sigma})$ has a universal profile near the cusps, which is independent of $\rho_0$, so that it is no longer sensitive to jumps in that parameter as we move across the junction points.

We now study $h_1 (\rho)$ in more detail:
\begin{eqnarray}
 h_1 (\rho_1) & = & 0 \nonumber \\
 \partial_\rho^k h_1 (\rho_1) & = & 2^{k-1} (1-\omega^2)
  \begin{cases}
   \sinh 2 \rho_1 & k \textrm{ even} \\
   \cosh 2 \rho_1 & k \textrm{ odd}
  \end{cases}
\label{eq:h1_and_derivs_at_rho_1_patched_K} \nonumber
\end{eqnarray}
It is easy to check that $w_1 = \cosh 2 \rho_1 = 1/\eta + O(1)$ and $\sinh 2 \rho_1 = 1/\eta + O(1)$, which then implies $\partial_\rho^k h_1 (\rho_1) = O(1)$, $\forall k > 0$. Of course, no discontinuities arise from this factor. Next, we consider the troublesome function $h_2(\rho)$:
\begin{eqnarray}
 && h_2 (\rho_1) = 1 - \frac{\sinh^2 2 \rho_0^{(j)}}{\sinh^2 2 \rho_1} \nonumber \\
 && \partial_\rho h_2 (\rho_1) = 2 \sinh^2 2 \rho_0^{(j)} \frac{\cosh 2 \rho_1}{\sinh^3 2 \rho_1} \nonumber \\
 && \partial_\rho^k h_2 (\rho_1) = 2^k \sinh^2 2 \rho_0^{(j)} \frac{P_k (2\rho_1)}{\sinh^{k+2} 2 \rho_1} \:, \qquad \textrm{for } \tilde{\sigma}_{j-1} \leq
  \tilde{\sigma} \leq \tilde{\sigma}_j 
\label{eq:h2_and_derivs_at_rho_1_patched_K} \nonumber
\end{eqnarray}
where $P_k (2 \rho_1)$ is a polynomial of degree $k$ in $\cosh 2\rho_1$ and $\sinh 2\rho_1$. It is now easy to deduce that $h_2 (\rho_1) \to 1$ and $\partial_\rho^k h_2 (\rho_1) \to 0$ as $\omega \to 1$, which means that $h_2 (\rho)$ becomes smooth at the junction points in this limit. Taking the behaviour of $h_1 (\rho)$ into account we can then deduce:
\begin{equation}
 \partial_\rho^k h (\rho_1) = \sum_{m=0}^k \partial_\rho^m h_1 (\rho_1) \partial_\rho^{k-m} h_2 (\rho_1) \to \partial_\rho^k h_1 (\rho_1) = O(1) \:, \qquad \textrm{as } \omega \to 1
\label{eq:behaviour_of_derivs_of_h_at_rho1_as_omega_to_1_patched_K} \nonumber
\end{equation} 
and therefore $h(\rho)$ also becomes smooth as $\omega \to 1$. This immediately shows that $\partial_{\tilde{\sigma}}^2 \rho (\tilde{\sigma})$ from \eqref{eq:rho''_patched_K} is continuous in the same limit. Now, working from that equation, we see that, in general, $\partial_{\tilde{\sigma}}^k \rho (\tilde{\sigma})$ is a sum of products of derivatives of $h (\rho)$, up to order $k-1$, and of derivatives of $\rho$, up to order $k-2$. The latter can all be re-expressed in terms of lower derivatives of $h (\rho)$ through \eqref{eq:rho''_patched_K} and \eqref{eq:def_h_h1_h2_studying_derivs_of_rho_patched_K}, so that, in the end, we're only left with derivatives of $h (\rho)$ which all become smooth in the limit considered (in particular, notice that there are never any diverging factors involved, so that the exponential suppression of the discontinuities as $\omega \to 1$ is never undone). Hence, $\partial_{\tilde{\sigma}} \rho (\tilde{\sigma})$ becomes a smooth function as $\omega \to 1$.

If we now consider $l(\rho)$, we easily see that:
\begin{equation}
 \partial_\rho^k l (\rho_1) = \frac{P_k (\rho_1)}{\sinh^{k+2} \rho_1} \to 0 \:, \qquad \textrm{as } \omega \to 1
\label{eq:l_and_derivs_patched_K} \nonumber
\end{equation}
which then implies that all derivatives of $g(\tilde{\sigma})$ at the junction points vanish as these points approach the boundary, since they are given by sums of products of derivatives of $l(\rho)$ and of $\rho(\tilde{\sigma})$ (these then reduce to derivatives of $h(\rho)$ through $(\partial_{\tilde{\sigma}} \rho )^2 = h(\rho)$ and $\partial_{\tilde{\sigma}}^2 \rho = h'(\rho)/2$); the former vanish, while the latter do not diverge. Thus, $g(\tilde{\sigma})$ and, similarly, $f(\tilde{\sigma})$ both become smooth as $\omega \to 1$.

Consequently, the patched Kruczenski string is an approximate solution which only becomes exact in the limit of large angular momentum, as the spikes approach the boundary of $AdS_3$.

It is also important to notice that the $\omega = 1$ solution \eqref{eq:JJ_single_arc_solution_omega=1}, representing a single arc with endpoints on the boundary, displays a different type of behaviour, since, by definition, it satisfies $\partial_{\tilde{\sigma}} \rho (\tilde{\sigma}) = h_2 (\rho)$, and thus now the first derivative of $\rho(\tilde{\sigma})$ no longer vanishes at the endpoints, where the spikes should be located. The reason is that in this solution we see the extreme consequences of the $\omega \to 1$ limit: the cusps are ``pushed away'' at infinity and eventually disappear from the worldsheet, which is now infinitely long. It is hence necessary to maintain $\omega > 1$ and then to study the type of limit we used just above, in order to keep track of the spikes.

\begin{figure}%
\includegraphics[width=\columnwidth]{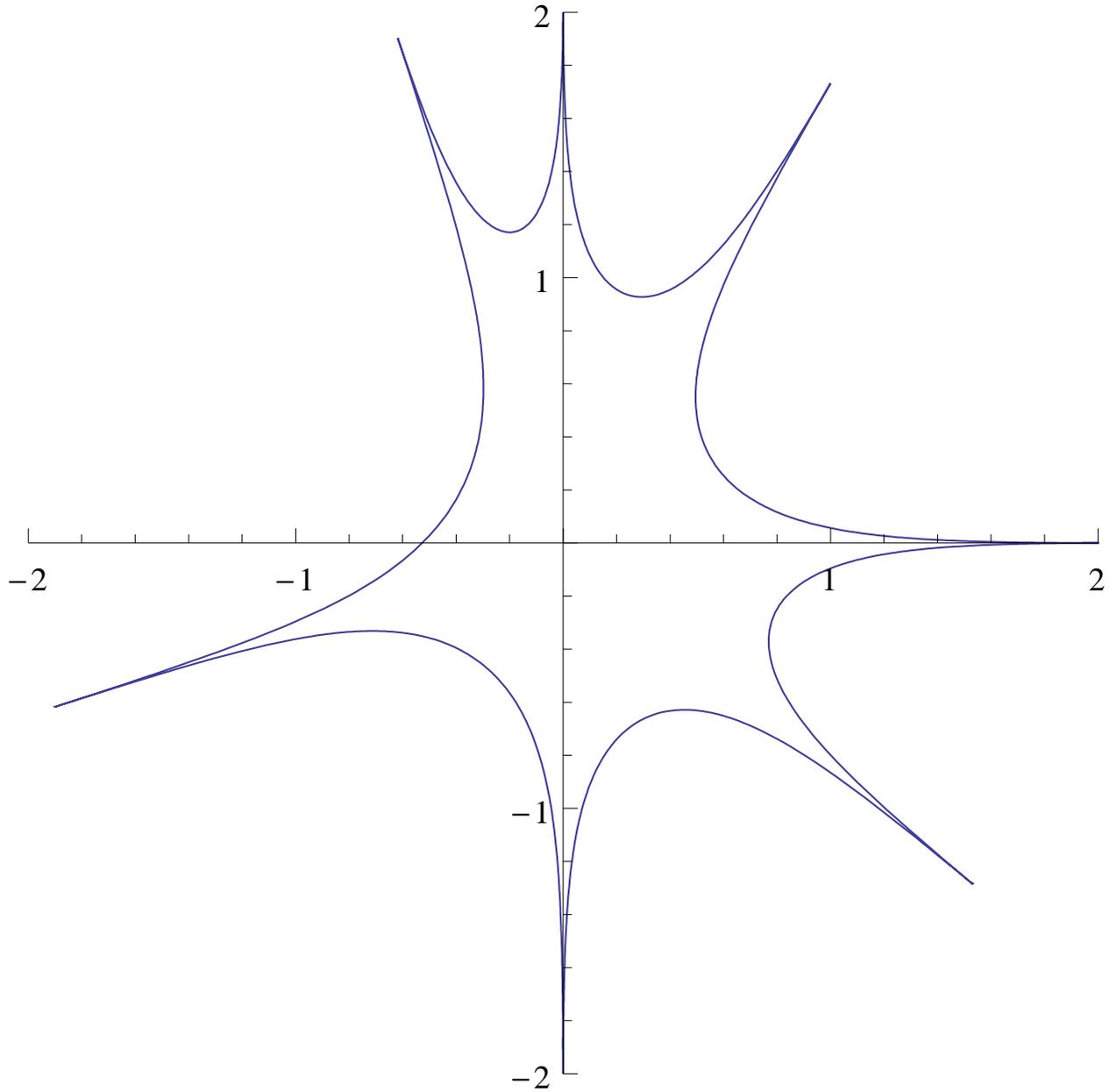}
\caption{The patched Kruczenski spiky string, with 7 spikes, angular separations equal to $\pi/3$, $\pi/6$, $\pi/10$, $\pi/2$, $2\pi/5$, $5\pi/18$, $2\pi/9$ and corresponding parameters given by $\rho_1 = 2$, $\rho_0^{(1)} = 0.638475$, $\rho_0^{(2)} = 0.965792$, $\rho_0^{(3)} = 1.18657$, $\rho_0^{(4)} = 0.43007$, $\rho_0^{(5)} = 0.546767$, $\rho_0^{(6)} = 0.727608$, $\rho_0^{(7)} = 0.833672$.}
\label{fig:patched_K_7_cusps}
\end{figure}

We now go back to the analysis of the general properties of the patched string. By construction, this solution, when plotted at constant $t$, has $K$ arcs of angular separation $\Delta\theta_j$, for $j=1,\ldots,K$, and hence the closedness condition is:
\begin{equation}
 \sum_{j=1}^K \Delta\theta_j = 2 n \pi
\label{eq:closedness_constraint_patched_K} \nonumber
\end{equation}
Clearly $\rho(\tilde{\sigma}_j)=\rho_1$, and thus we have $K$ cusps located at $\tilde{\sigma} = \sigma_j$, for $j=0,\ldots,K-1$ (the analysis of section \ref{sec:general_properties_Kruc} still applies, and thus we have $\partial_{\tilde{\sigma}} (t,\rho,\phi) = (0,0,0)$ at each cusp). We denote their angular positions by $\phi_j \equiv \omega \tilde{\tau} + \theta_j$, where, without loss of generality, we can assume $\theta_0 = 0$. We also define $\theta_{K} \equiv 2 n \pi - \theta_0 = 2 n \pi$, so that $\Delta\theta_j = \theta_j - \theta_{j-1}$ and $\sum_{j=1}^m \Delta\theta_j = \theta_m$. The plot at constant time $t$ is shown in Fig. \ref{fig:patched_K_7_cusps}.

As usual, we will be interested in the limit of this solution as the spikes touch the boundary of $AdS_3$, i.e. $\omega\to 1$, in which $\rho_1 \to + \infty$ and $\rho_0^{(j)}$ approaches the value which satisfies the following equation, coming from \eqref{eq:Deltatheta_single_arc_JJ_omega_to_1}:
\begin{equation}
 \Delta\theta_j \simeq 2 \mathrm{Arctan} \frac{1}{\sinh 2\rho_0^{(j)}} \:, \qquad \textrm{as } \omega \to 1
\label{eq:Deltatheta_single_arc_patched_K_omega_to_1} \nonumber
\end{equation}

\subsection{Energy, angular momentum and large angular momentum behaviour}
\label{sec:patched_K_E_S_large_S_behaviour}

The energy and angular momentum are computed from \eqref{eq:def_patched_K_solution}, \eqref{eq:AdS3_energy} and \eqref{eq:AdS3_spin}. They clearly reduce to the sum of the usual Kruczenski-type contributions from each individual arc:
\begin{eqnarray}
 \Delta 
   & = & \frac{\sqrt{\lambda}}{2\pi} \sum_{j=1}^K \int_0^{2\tilde{L}_j} d\tilde{\sigma} \cosh^2 \rho
          (\tilde{\sigma},\rho_0^{(j)}) \nonumber \\
   & = & \sum_{j=1}^K \frac{\sqrt{\lambda}}{\pi} \sqrt{\frac{w_1-1}{w_1+w_0^{(j)}}} \left[ \frac{1}{2} (w_1+w_0^{(j)}) \mathbb{E}_j - 
          \sinh^2 \rho_0^{(j)} \mathbb{K}_j \right]
   \label{eq:E_patched_K}\nonumber\\
 S 
   & = & \frac{\omega\sqrt{\lambda}}{2\pi} \sum_{j=1}^K \int_0^{2\tilde{L}_j} d\tilde{\sigma} \sinh^2 \rho
          (\tilde{\sigma},\rho_0^{(j)}) \nonumber \\
   & = & \sum_{j=1}^K \frac{\omega\sqrt{\lambda}}{\pi} \sqrt{\frac{w_1-1}{w_1+w_0^{(j)}}} \left[ \frac{1}{2} (w_1+w_0^{(j)}) \mathbb{E}_j - 
          \cosh^2 \rho_0^{(j)} \mathbb{K}_j \right] \nonumber \\
   \label{eq:S_patched_K}\nonumber
\end{eqnarray}
where $\rho(\tilde{\sigma},\rho_0)$ again represents the original solution \eqref{eq:rho_of_sigmatilde_JJ}.

As $\omega\to 1$, with $\omega = 1+\eta$, we have:
\begin{eqnarray}
 \Delta & = & K \frac{\sqrt{\lambda}}{2 \pi \eta} - \frac{\sqrt{\lambda}}{32\pi} \sum_{j=1}^K (8+3w_0^{(j)}) \log \eta \nonumber \\
   &   & + \frac{\sqrt{\lambda}}{64\pi} \sum_{j=1}^K \left[ -13w_0^{(j)} + (32+12w_0^{(j)}) \log \frac{2 \sqrt{2}}{\sqrt{w_0^{(j)}}} \right] 
          + O(\eta\log\eta)
           \label{eq:behaviour_of_E_patched_K_as_omega_goes_to_1}\nonumber \\
 S & = & K \frac{\sqrt{\lambda}}{2 \pi \eta} + \frac{\sqrt{\lambda}}{32\pi} \sum_{j=1}^K (8-3w_0^{(j)}) \log \eta \nonumber \\
   &   & + \frac{\sqrt{\lambda}}{64\pi} \sum_{j=1}^K \left[32 -13w_0^{(j)} + (-32+12w_0^{(j)}) \log \frac{2 \sqrt{2}}{\sqrt{w_0^{(j)}}} \right]
          + O(\eta\log\eta)\nonumber \\
           \label{eq:behaviour_of_S_patched_K_as_omega_goes_to_1}
\end{eqnarray}
and we can, as usual, compute the $O(1)$ correction to the typical logarithmic growth of the anomalous dimension:
\begin{equation}
 \Delta - S = \frac{K\sqrt{\lambda}}{2\pi} \log \left( \frac{2\pi S}{K\sqrt{\lambda}} \right) + \frac{\sqrt{\lambda}}{2\pi} \left[ K (3\log 2 - 1) +
  \sum_{j=1}^K \log \left( \sin \frac{\Delta\theta_j}{2} \right) \right] + O(\eta\log\eta)
\label{eq:E-S_omega_to_1_behaviour_patched_K}
\end{equation}
where we have used:
\begin{equation}
 \frac{1}{w_0^{(j)}} \simeq \sin \frac{\Delta\theta_j}{2} \qquad \textrm{as } \omega \to 1
\label{eq:relation_between_u0j_and_sin_Deltathetaj/2} \nonumber
\end{equation}
which generalises \eqref{eq:relation_between_u0_and_sin_Deltatheta/2}.

\subsection{Spectral curve for large S}
\label{sec:spectral_curve_patched_K}

As always, we introduce the rescaled worldsheet coordinates, defined as:
\begin{equation}
 (\tau,\sigma) = \frac{2\pi}{L} (\tilde{\tau},\tilde{\sigma}) = \frac{\pi}{\sum_{j=1}^K \tilde{L}_j}(\tilde{\tau},\tilde{\sigma})
\label{eq:def_rescaled_coords_patched_K} \nonumber
\end{equation}
in terms of which the cusp positions become:
\begin{equation}
 \sigma_j = \frac{4\pi}{L} \sum_{k=1}^j \tilde{L}_k = 2\pi \frac{\sum_{k=1}^j \tilde{L}_k}{\sum_{k=1}^K \tilde{L}_k} \neq \frac{2j\pi}{K}
   \:, \qquad \textrm{for } j=0,\ldots,n-1
\label{eq:cusp_positions_patched_K_sigma}
\end{equation}
We then substitute \eqref{eq:def_patched_K_solution} into \eqref{eq:components_of_j_tau_in_terms_of_t_phi_rho} and write everything in terms of $(\tau,\sigma)$, thus obtaining the rescaled charge density:
\begin{eqnarray}
 j^0_\tau(\tau,\sigma) & = & \frac{L}{2\pi} \left\{ (\omega+1) \left[ w_1 \mathrm{cn}^2 \left(
  \left. v^{(j)} \left(\frac{L}{2\pi}(\sigma-\sigma_{j-1})\right) \right| k^{(j)} \right) \right. \right. \nonumber \\
  & & \left. + w_0^{(j)} \left. \mathrm{sn}^2 \left( \left. v^{(j)}\left(\frac{L}{2\pi}(\sigma-\sigma_{j-1})\right) \right|k^{(j)} \right)
   \right] + 1-\omega  \right\} \nonumber \\
 j^1_\tau(\tau,\sigma) + i j^2_\tau(\tau,\sigma) & = & i \frac{L}{2\pi} (\omega+1) e^{i(\phi-t)} \nonumber \\
  & & \times \left\{ \left[ w_1 \mathrm{cn}^2 \left( \left. v^{(j)}\left(\frac{L}{2\pi}(\sigma-\sigma_{j-1})\right) \right|k^{(j)} \right)
   \right. \right. \nonumber \\
  & & + \left. \left. w_0^{(j)} \mathrm{sn}^2 \left( \left. v^{(j)}\left(\frac{L}{2\pi}(\sigma-\sigma_{j-1})\right) \right|k^{(j)} \right)
   \right]^2 - 1\right\}^\frac{1}{2}\nonumber \\
  & & \qquad\qquad\qquad\qquad\qquad\qquad \textrm{for } \sigma \in [\sigma_{j-1},\sigma_j]
\label{eq:eq:j_tau_in_terms_of_tau_sigma_patched_K}
\end{eqnarray}
As before, we express $\sigma$ near the $m$-th spike as $\sigma = \sigma_m + \hat{\sigma}$, where $\hat{\sigma}$ is never allowed to reach one half of the distance to the nearest cusp in both directions. For the patched Kruczenski solution, this translates into an asymmetric condition on $\hat{\sigma}$ (since the fundamental periods $\tilde{L}_j$ are in general different from each other), which, however, reduces to the usual $|\hat{\sigma}| \leq \pi/K$ as $\omega \to 1$ (since all $\tilde{L}_j$ become identical in this limit).

Then, it is only a matter of tedious algebra to carry out the usual expansion of the elliptic functions and integrals as $\omega \to 1$ and obtain:
\begin{eqnarray}
 j_\tau^0 (\tau,\sigma) & \simeq & \frac{2 \mathbb{K}}{\pi} \frac{1}{\eta \cosh^2 \left( \frac{\mathbb{K}}{\pi} \hat{\sigma} \right)} \nonumber\\
 j^1_\tau(\tau,\sigma) + i j^2_\tau(\tau,\sigma) & \simeq & i \frac{2 \mathbb{K}}{\pi} \frac{1}{\eta \cosh^2 \left( \frac{\mathbb{K}}{\pi}
  \hat{\sigma} \right)} e^{i \sum_{j=1}^m \Delta\theta_j} 
\label{eq:leading_behaviour_of_jtau_patched_K} \nonumber
\end{eqnarray}
where $\mathbb{K} \equiv \sum_{j=1}^K \mathbb{K}_j$. As in the previous cases, we can use this result to compute the normalised charge density:
\begin{eqnarray}
 \mu^0 (\tau,\sigma) & = & \frac{1}{K} \sum_{m=0}^{K-1} \delta (\sigma - \sigma_m) \nonumber\\
 \mu^1 (\tau,\sigma) + i \mu^2 (\tau,\sigma) & = & i \frac{1}{K} \sum_{m=0}^{K-1} e^{i \sum_{j=1}^m \Delta\theta_j} \delta (\sigma - \sigma_m)
\label{eq:eq:cpts_of_mu_patched_K}
\end{eqnarray}
The corresponding spin vectors are:
\begin{equation}
 \vec{L}_m = \frac{S}{K} \begin{pmatrix}
	                  1 \\
	                  -\sin \left( \sum_{j=1}^m \Delta\theta_j \right) \\
	                  \cos \left( \sum_{j=1}^m \Delta\theta_j \right)
                   \end{pmatrix} = \frac{S}{K} \begin{pmatrix}
	                                                   1 \\
	                                                   -\sin \theta_m \\
	                                                   \cos \theta_m
                                                    \end{pmatrix}
\label{eq:spin_vector_at_mth_cusp_patched_K}
\end{equation}
(we recall that $\theta_0 = 0$). These vectors satisfy the second property listed in \eqref{eq:properties_of_spin_vectors}, but not the first, i.e. the highest weight condition (in particular, the integrals over $\sigma$ of $\mu^1$ and $\mu^2$ don't vanish, whereas the integral of $\mu^0$ is still equal to 1). It is however possible to obtain the highest weight state by performing a right $SU(1,1)$ rotation of the patched Kruczenski solution, as described in \eqref{eq:SU11_left_and_right_multiplication}. It is easy to see from the definition that the right current $j$ transforms as: $j_a \to j'_a = U_R^{-1} j_a U_R$. In general, such a rotation can modify the first component of $j_a$, and it is possible to show that the type of rotation required in order to find the highest weight state will do so, and thus $\Delta'+S'\neq \Delta+S$. On the contrary, the left current $l$ is invariant under this rotation, and hence we have $\Delta'-S'=\Delta-S$, which implies both $\Delta'\neq \Delta$ and $S'\neq S$. Consequently, a different spin $S'$ appears in the definition \eqref{eq:def_normalised_charge_density} of the new normalised charge density $\mu'$, which can be expressed in terms of $\mu$ as follows:
\begin{equation}
 \mu'(\tau,\sigma) = U_R^{-1} \mu(\tau,\sigma) U_R \lim_{\omega\to 1} \frac{S}{S'}
\label{eq:muprime_in_terms_of_mu_R_SU11_rotation} \nonumber
\end{equation}
We can now impose that the integrals of $\mu'^1$ and $\mu'^2$ over $\sigma$ vanish (the integral of $\mu'^0$ is automatically 1 due to the new normalisation factor, which contains $S'$ instead of $S$) and solve for the parameters of the rotation. As a consequence of their definition \eqref{eq:def_spin_vector_at_each_cusp}, the spin vectors (which are $\mathfrak{su}(1,1)$ matrices), transform as:
\begin{equation}
 L_m' = U_R^{-1} L_m U_R \nonumber
\label{eq:transf_of_L_m_under_right_rotation}
\end{equation}
and it is then easy to see from the definition $\mathbb{L}_m (u) = \mathbb{I} u + \eta_{AB} L_m^A s^B /S$ that the same applies to the matrix $\mathbb{L}_m (u)$:
\begin{equation}
 \mathbb{L}'_m (u) = U_R^{-1} \mathbb{L}_m (u) U_R
\label{eq:L_m(utilde)_prime_in_terms_of_L_m(utilde)}
\end{equation}
It is now clear from this last equation that the trace of the monodromy matrix, as a function of the variable $u$, is also invariant, and consequently the same is true of the spectral curve and of all the conserved charges $q_k$ associated with the spin chain. Therefore, rather than explicitly performing a $\{ \Delta\theta_j \}$-dependent $SU(1,1)$ rotation, we'll continue working with the current version of the patched Kruczenski spiky string, and all results will equally apply to the case of the highest weight state.

We can now proceed along the usual path and calculate the matrix $\mathbb{L}_m (u)$ from \eqref{eq:spin_vector_at_mth_cusp_patched_K}:
\begin{equation}
 \mathbb{L}_m (u) = \begin{pmatrix}
	                           u + \frac{i S}{K} & \frac{i S}{K} e^{i \theta_m} \\
	                           - \frac{i S}{K} e^{-i \theta_m} & u - \frac{i S}{K}
                            \end{pmatrix} %\equiv \begin{pmatrix}
	                                        %        a & b_m \\
	                                        %        \bar{b}_m & \bar{a}
                                          %       \end{pmatrix}
\label{eq:L_m(utilde)_for_patched_K} \nonumber
\end{equation}
As expected, this matrix coincides with the one computed for the
original Kruczenski solution\footnote{In order to have a perfect match
  with equation \eqref{eq:L_m(utilde)_for_JJ_1}, we could introduce an
  arbitrary shift $\phi_0$ on the angular coordinate, without changing
  anything in the previous discussion. However, by explicitly
  performing the calculation, it is easy to see that $\phi_0$ would
  quickly disappear from the result anyway, as it did in the previous
  cases.} if we choose $\Delta\theta_j = 2 n \pi/K$, $\forall
j$, and therefore, under this condition, all subsequent results will 
reduce to those we obtained for that solution.

Due to the arbitrariness of the angular separations $\Delta\theta_j$, the procedure we used in section \ref{sec:spectral_curve_JJ} in order to calculate $\mathrm{tr} \: \Omega[x]$ is no longer effective. Nonetheless, it is possible, through a tedious calculation which however only involves elementary reasoning, to show that the conserved charges $q_k$ (we recall that $\hat{\mathbb{P}}_K (1/u) = \mathrm{tr} \: \Omega[x]$ is related to these by \eqref{eq:general_curve_from_spin_chain}) can be expressed as:
\begin{multline}
 q_k = 2 \left( \frac{S}{K} \right)^k \sum_{r=0}^{\left[ \frac{k}{2} \right]} (-1)^r \sum_{\begin{array}{l}
	                                                                                 d_1,\ldots,d_r = 0,\ldots,K-2r \\
	                                                                                 D \leq K-2r
                                                                                  \end{array}} C(k,K,r,D)\\
    \times \sum_{\begin{array}{l}
	          j_1,\ldots,j_r = 2,\ldots,K \\
	          j_{l+1} > j_l + d_l + 1
           \end{array}}
    \mathrm{Re} \left[ i^k e^{-i \sum_{l=1}^r (\theta_{j_l+d_l} - \theta_{j_l - 1})} \right]
\label{eq:qtilde_k_from_trace_patched_K}
\end{multline}
where we define:
\begin{equation}
 \begin{gathered}
  D \equiv \sum_{l=1}^r d_l \:, \\
  C(k,K,r,D) \equiv \sum_{j=\max\{k-K+D,0\},\ldots,\min\{D,k-2r\}} (-1)^j \binom{K-2r-D}{k-2r-j} \binom{D}{j} 
 \end{gathered}
\label{eq:def_of_D_Deltatheta_jd_and_C(knrD)_patched_K} \nonumber
\end{equation}
Since the method we used in obtaining \eqref{eq:qtilde_k_from_trace_patched_K} defines $q_k$ as the coefficient of $1/u^k$ in $\mathrm{tr} \: \Omega[x]$, this expression also correctly reproduces the known term, $q_0 = 2$, and the missing linear term in $1/u$, $q_1 = 0$.

A less unwieldy expression for $q_k$ can be obtained by identifying
the string theory equivalents of the 
spin chain variables $z_k$ and $p_k$ introduced
in \cite{KK2}, for $k =
1, \ldots, K$ (where $K$ is the number of spins in the chain, which
matches the number of cusps). As described
in \cite{KK2}\footnote{In their notation, $L_3 \equiv \sum_{k=1}^N
  \mathcal{L}_k^0$, $L_+ \equiv \sum_{k=1}^N \mathcal{L}_k^+$ and $L_-
  \equiv \sum_{k=1}^N \mathcal{L}_k^-$.}, these parameters are related
to the individual spin vectors 
$\mathcal{L}_k$ at each site of the chain:
\begin{equation}
 \mathcal{L}_k^0 = i z_k p_k \:, \quad \mathcal{L}_k^+ = i z_k^2 p_k \:, \quad \mathcal{L}_k^- = -i p_k
\label{eq:spin_chain_spin_vectors_at_each_site_in_terms_of_zk,pk_patched_K} \nonumber
\end{equation}
According to \cite{D1}, we relate these to the spin vectors at each cusp \eqref{eq:spin_vector_at_mth_cusp_patched_K} as follows:
\begin{equation}
 L_k^0 = \mathcal{L}_k^0 \:, \qquad i \mathcal{L}_k^\pm = L_k^1 \pm i L_k^2
\label{eq:relation_between_spin_vect_at_each_site_and_spin_vect_at_each_cusp_patched_K} \nonumber
\end{equation}
It is now straightforward to obtain:
\begin{equation}
 p_k = - \frac{i S}{K} e^{-i \theta_k} \:, \qquad z_k = e^{i \theta_k}
\label{eq:identification_of_spin_chain_parameters_in_stringy_calculation_patched_K} \nonumber
\end{equation}
Then, the $k$-th conserved charge is given by:
\begin{equation}
 q_k = \sum_{1\leq j_1 < j_2 < \ldots < j_k \leq K} z_{j_1 j_2} z_{j_2 j_3} \ldots z_{j_{k-1} j_k} z_{j_k j_1} p_{j_1} p_{j_2} \ldots p_{j_k}
  \:, \quad k=2, \ldots, K
\label{eq:q_k_from_spin_chain_patched_K}
\end{equation}
where $z_{ab} = z_a - z_b$. After some algebra, we can recast this expression into the following form:
\begin{equation}
 q_k =  \left( - \frac{2 S}{K} \right)^k \sum_{1\leq j_1 < j_2 < \ldots < j_k \leq K} \prod_{l=1}^k \sin \left(
  \frac{\theta_{j_{l+1}} - \theta_{j_l}}{2} \right) \:, \quad k=2, \ldots, K
\label{eq:qktilde_from_spin_chain_patched_K}
\end{equation}
where we have defined $j_{k+1} \equiv j_1$. A tedious but straightforward calculation shows that this expression equals \eqref{eq:qtilde_k_from_trace_patched_K} for $k\geq 2$, and thus yields the conserved charges from $\mathrm{tr} \: \Omega$. By using this result, we show, in appendix \ref{sec:app_computing_qtilde_2_for_N-folded_GKP_and_K}, that $q_2$ has a complicated dependence on the angular separations $\Delta \theta_j$ and that, in general, $q_2 \neq - S^2$, unless all the $\Delta\theta_j$ are equal, which, as we can see from \eqref{eq:eq:cpts_of_mu_patched_K}, is precisely the condition for the patched solution to be a highest weight state. This is exactly the behaviour we expected from the general finite gap picture: $q_2$ only equals $-S^2$ when the string considered is a highest weight state.

As usual, we're interested in the highest conserved charge, which is given by:
\begin{eqnarray}
 q_K & = & \left( - \frac{2 S}{K} \right)^K \prod_{l=1}^K \sin \left( \frac{\theta_{l+1} - \theta_l}{2} \right) \nonumber\\
             & = & \left( - \frac{2 S}{K} \right)^K \sin \left( \frac{\Delta\theta_1 - 2 n \pi}{2} \right) \prod_{l=1}^{K-1} \sin \left(
              \frac{\Delta\theta_{l+1}}{2} \right) \nonumber \\
             & = & \left( - \frac{2 S}{K} \right)^K (-1)^{n} \prod_{l=1}^{K} \sin \left( \frac{\Delta\theta_{l}}{2} \right)
\label{eq:qtilde_n_for_patched_K}
\end{eqnarray}
where we have used $\theta_1 - \theta_K = - \sum_{m=2}^K \Delta\theta_m = - 2 n \pi + \Delta\theta_1$ in deriving the second line. We can now substitute this into \eqref{eq:general_gauge_theory_prediction_for_E-S} in order to obtain the gauge theory prediction for $\Delta-S$:
\begin{eqnarray}
 \Delta - S & = & \frac{K\sqrt{\lambda}}{2\pi} \log S + \frac{\sqrt{\lambda}}{2\pi} \left[ K \log 2 - K \log K + \log (-1)^{K + n}
  + \sum_{j=1}^K \log \left( \sin \frac{\Delta\theta_j}{2} \right) \right. \nonumber\\
 & & \left. \phantom{\log \left( \sin \frac{\Delta\theta_j}{2} \right)} + C_{\mathrm{string}}(K) \right]
\label{eq:gauge_theory_prediction_for_E-S_patched_K} \nonumber
\end{eqnarray}
(where, as usual, we have omitted subleading terms as $\omega \to 1$) which, by comparison with \eqref{eq:E-S_omega_to_1_behaviour_patched_K}, implies:
\begin{equation}
 C_{\mathrm{string}}(K) = K \left[ \log \left( \frac{8 \pi}{\sqrt{\lambda}} \right) - 1 \right] - \log (-1)^{K + n}
\label{eq:C(n)_from_patched_K} \nonumber
\end{equation}
This expression agrees with those obtained in the previous two cases, \eqref{eq:C(n)_from_JJ} and \eqref{eq:C(n)_from_Nfolded_GKP} (in order to obtain the latter, we must set $K$ even and $n = K/2$, as we remarked at the end of section \ref{sec:spectral_curve_JJ}).

Finally, we'd like to observe that, as we noticed earlier in this section, it is possible to obtain all the previous results concerning the N-folded GKP string and the Kruczenski string from this generalised version. For instance, it is easy to check that the highest conserved charge \eqref{eq:qtilde_n_for_patched_K} reduces to the expression \eqref{eq:highest_conserved_charge_JJ} valid in the Kruczenski case if we set $\Delta\theta_j = 2 n \pi /K$, $\forall j$.

\section{Results and Interpretation}

In this paper we have constructed explicit string solutions with
large angular momentum whose spectral curves and energies agree
precisely with those of the classical spin chain arising in the dual 
gauge theory. This is consistent with the exact equality of the
semiclassical spectra of the two theories observed in \cite{D1}. In
principle one can reconstruct the gauge theory operator corresponding
to any large-$S$ string solution. In particular the filling fractions for Bethe
roots, which define an eigenstate of the dilatation operator at
one-loop can be extracted from the curve 
using the semiclassical formulae (\ref{BS1}). 
   
The results reported in this paper are consistent with the explicit
identification between string theory and gauge theory proposed in
\cite{D1}. For all the solutions constructed above, the large-$S$ limit
of the angular momentum density $j_{\tau}$ is $\delta$-function
localised at the spikes and has the form, 
\bea 
j_{\tau}(\sigma,\tau) & \sim & 
\frac{8\pi}{\sqrt{\lambda}} \sum_{k=0}^{K-1} L_k \delta(\sigma -
\sigma_k) \nn 
\eea
for some $\mathfrak{su}(1,1)$-valued variables $L_{k}$. The
proposal of \cite{D1} was that these variables should be identified
directly with the classical spin variables of the one-loop spin chain
introduced in Section 2 according to $\mathcal{L}^{0}_{k}=
L^{0}_{k}$ and $i\mathcal{L}^{\pm}_{k}=L^{1}_{k}\pm iL^{2}_{k}$. 
As explained in Section 6 of \cite{D1}, this identification implies 
the agreement of the gauge theory and string theory curves. Conversely 
the agreement of the curves found above provides a non-trivial test of
the proposed identification. 
In the previous section we also obtained an explicit formula 
(\ref{eq:spin_vector_at_mth_cusp_patched_K}) for the variables $L_{k}$
evaluated on the general multi-spike solution.  
\paragraph{}
An interesting perspective on the results of this paper is obtained by
invoking the standard correspondence between local operators of the
${\cal N}=4$ theory in Minkowski space and states of the same theory
defined on a three-sphere. In particular the operators (\ref{sl2})
correspond to states comprising of $J$ massless quanta of the adjoint
scalar field with angular momenta $s_{j}$, $j=1,2,\ldots J$, on
$S^{3}$. In the large spin limit, $K\leq J$ particles have large
angular momentum and therefore can be assigned 
to classical orbits around a great circle on $S^{3}$. Each particle is a
source for chromoelectric flux, and successive particles are joined by
flux lines which spread out on the sphere reflecting the non-abelian
Coulomb phase of the ${\cal N}=4$ theory. It is interesting to compare
this state with the corresponding solution in string theory on $S^{3}$ 
which corresponds to a string with $K$ cusps approaching the
boundary. The cusps correspond to localised regions of energy/angular
momentum density in the boundary theory while the arcs of string 
drooping into the interior of $AdS_{3}$ correspond in the usual way 
to flux spreading out on the boundary. The two pictures are in
surprisingly good agreement. One interesting feature of this agreement
is the correspondence between cusps and the massless adjoint quanta
of the field theory \cite{AlMal}. At least for large spin, it seems that the
partonic nature of the 
gauge-invariant state created by the operators (\ref{sl2}) persists 
at strong coupling. 
\paragraph{}
Finally there are several interesting extensions of this work that
could be studied. The first is to consider more general
operators of the ${\cal N}=4$ theory including different covariant
derivatives and scalars as well as fermions. This requires a treatment
of the string theory on $AdS_{5}\times S^{5}$. Another interesting
direction would be to quantize the dynamics of the spikes. Finally, it
would be interesting to consider the semi-classical limit of the
large-$N$ QCD spin chain of \cite{BFZ} and try to interpret it in
terms of a semiclassical string theory.   
\paragraph{}
ND would like to thank Gregory Korchemsky and Arkady Tseytlin for
useful discussions. 

\appendix

\section{Gauge transformation for the Kruczenski solution}
\label{sec:app_gauge_transf_for_K}

The Kruczenski spiky string is described by the following ansatz: $t = \tau$, $\rho = \rho(\sigma)$, $\phi = \omega \tau + \sigma$, which guarantees that all equations of motion from the Nambu-Goto action are satisfied if $\rho(\sigma)$ solves the following:
\begin{equation}
 \rho' = \pm \frac{1}{2} \frac{\sinh 2\rho}{\sinh 2\rho_0} \frac{\sqrt{\sinh^2 2\rho - \sinh^2 2\rho_0}}{\sqrt{\cosh^2 \rho - \omega^2 \sinh^2
  \rho}}
\label{eq:Nambu-Goto_eom_Kruc}
\end{equation}
where $\rho_0$ is an integration constant. The requirement of reality placed upon $\rho$ forces $\rho_0 \leq \rho \leq \rho_1$, where $\coth \rho_1 = \omega$. From now on, we will refer to the function which solves equation \eqref{eq:Nambu-Goto_eom_Kruc} as $\hat{\rho}(\sigma)$. It is possible to integrate \eqref{eq:Nambu-Goto_eom_Kruc} to get the inverse function $\sigma(\hat{\rho})$:
\begin{equation}
 \sigma = \pm \frac{\sinh 2\rho_0}{\sqrt{2}\sqrt{w_0 + w_1} \sinh\rho_1} \left\{ \Pi \left( \frac{w_1-w_0}{w_1-1}, \beta, p \right) - \Pi \left(
  \frac{w_1-w_0}{w_1+1}, \beta, p \right) \right\}
\label{eq:sigma_of_rho_Kruc_Nambu-Goto}
\end{equation}
where:
\begin{equation}
 p \equiv \sqrt{\frac{w_1-w_0}{w_1+w_0}} \:, \qquad \sin\beta \equiv \sqrt{\frac{w_1-w(\hat{\rho})}{w_1-w_0}}
\label{eq:p_and_beta_in_Kruc_Nambu-Goto}
\end{equation}
($\beta \in [0,\pi/2]$) and we define $w(x) \equiv \cosh (2x)$, $w_0 \equiv \cosh 2\rho_0$ and $w_1 \equiv \cosh 2\rho_1$.

We can construct a spiky string from this object by taking \eqref{eq:sigma_of_rho_Kruc_Nambu-Goto} with the plus sign and then replacing $\beta$ with the new coordinate $\sigma'$:
\begin{equation}
 \sigma = \frac{\sinh 2\rho_0}{\sqrt{2}\sqrt{w_0 + w_1} \sinh\rho_1} \left\{ \Pi \left(\frac{w_1-w_0}{w_1-1}, \sigma', p \right) - \Pi \left(
  \frac{w_1-w_0}{w_1+1}, \sigma', p \right) \right\}
\label{eq:def_sigmaprime_K}
\end{equation}
While \eqref{eq:def_sigmaprime_K} implies that $\sigma$ is an increasing function of $\sigma'$ (with $\sigma(\sigma'=0)=0$), \eqref{eq:p_and_beta_in_Kruc_Nambu-Goto} allows us to express $\hat{\rho}$ as a function of $\sigma'$:
\begin{equation}
 \sinh^2 \hat{\rho} = \sinh^2 \rho_1 \cos^2 \sigma' + \sinh^2 \rho_0 \sin^2 \sigma'
\label{eq:rhohat_of_sigmaprime_K} \nonumber
\end{equation}
From \eqref{eq:def_sigmaprime_K}, we see that, for each increase of $\pi/2$ in $\sigma'$, $\sigma$ and consequently $\phi$ increase by:
\begin{equation}
 \Delta\phi = \frac{\sinh 2\rho_0}{\sqrt{2}\sqrt{w_0 + w_1} \sinh\rho_1} \left\{ \Pi \left( \frac{w_1-w_0}{w_1-1},p \right) - \Pi \left(
  \frac{w_1-w_0}{w_1+1},p \right) \right\}
\label{eq:def_deltaphi_K}
\end{equation}
Thus, for the string to be closed at fixed $t=\tau$, we allow $\sigma'$ to vary in $[0,K \pi]$, $K \in \mathbb{N}$ (i.e. $\sigma \in [0,2 K \Delta\phi]$), and then demand that the corresponding total increase in $\phi$ be an integer multiple of $2\pi$: $2 K \Delta\phi = 2n\pi$. Since \eqref{eq:def_deltaphi_K} matches \eqref{eq:deltaphi_for_JJ}, the closedness conditions for these two solutions are actually the same.

We are now ready to discuss the worldsheet coordinate transformation which maps this solution onto the corresponding conformal gauge version \eqref{eq:rho_of_sigmatilde_JJ}, \eqref{eq:f_and_g_of_sigmatilde_JJ}. In order to find it, we just need to impose the equality of the global coordinates $(t,\rho,\phi)$ specified by the two different versions of the ansatz, which leads to the following set of relations:
\begin{equation}
 \tilde{\tau} + f(\tilde{\sigma}) = \tau \:, \qquad g(\tilde{\sigma}) - \omega f(\tilde{\sigma}) = \sigma \:, \qquad
   \rho(\tilde{\sigma}) = \hat{\rho}(\sigma)
\label{eq:equality_of_global_coords_in_two_versions_of_ansatz_K_JJ}
\end{equation}
These are actually three conditions on two unknown functions $\tau(\tilde{\tau},\tilde{\sigma})$, $\sigma(\tilde{\tau},\tilde{\sigma})$, and we easily see that they give two potentially conflicting expressions for $\sigma(\tilde{\tau},\tilde{\sigma})$. For the transformation to exist, these must coincide:
\begin{equation}
 g(\tilde{\sigma}) - \omega f(\tilde{\sigma}) = \hat{\rho}^{-1}(\rho(\tilde{\sigma}))
\label{eq:compatibility_for_coord_transform_K_JJ} \nonumber \\
\end{equation}
We already have the inverse of $\hat{\rho}$ from \eqref{eq:sigma_of_rho_Kruc_Nambu-Goto}. We can then compute $w(\rho(\tilde{\sigma})) = \cosh 2 \rho(\tilde{\sigma})$ from \eqref{eq:rho_of_sigmatilde_JJ} and then use it to find:
\begin{equation}
 \sin^2 \sigma' = \frac{w_1 - w(\rho(\tilde{\sigma}))}{w_1 - w_0} = \mathrm{sn}^2(v|k)
\label{eq:sin_beta_checking_compatibility_of_coord_transform_K_JJ} \nonumber
\end{equation}
Remembering that $\sigma' \in [0,K \pi]$, it is natural to identify $\sigma' = \mathrm{am}(v|k)$. Therefore, by substituting this into \eqref{eq:def_sigmaprime_K}, we get:
\begin{equation}
 \hat{\rho}^{-1} (\rho(\tilde{\sigma})) = \frac{\sinh 2\rho_0}{\sqrt{2} \sinh \rho_1 \sqrt{w_1+w_0}} \left\{ \Pi (n_-, \mathrm{am} (v|k), k) -
  \Pi (n_+, \mathrm{am} (v|k), k) \right\}
\label{eq:sigma_from_rhohatinverse_and_rho} \nonumber
\end{equation}
It is now only a matter of simple algebra to show that this expression matches $g(\tilde{\sigma}) - \omega f(\tilde{\sigma})$, i.e. that the last two conditions in \eqref{eq:equality_of_global_coords_in_two_versions_of_ansatz_K_JJ} are equivalent, and thus that the coordinate transformation exists. Its explicit form is the following:
\begin{eqnarray}
 \tau & = & \tilde{\tau} + \frac{\sqrt{2}\omega\sinh 2\rho_0 \sinh \rho_1}{(w_1+1) \sqrt{w_0+w_1}} \Pi (n_+, \mathrm{am} (v|k), k) \nonumber \\
 \sigma & = & \frac{\sinh 2\rho_0 }{\sqrt{2} \sqrt{w_0+w_1} \sinh \rho_1} \left\{ \Pi (n_-, \mathrm{am} (v|k), k) - \Pi (n_+, \mathrm{am} (v|k),
  k) \right\}
\label{eq:coord_transf_between_NG_and_conf_gauge_K}
\end{eqnarray}

It is also possible to determine the worldsheet metric $h_{ab}$ from Kruczenski's parametrization and then show that \eqref{eq:coord_transf_between_NG_and_conf_gauge_K} brings it to the 2-dimensional Minkowski metric, up to a conformal transformation. We recall that the Nambu-Goto action is obtained from the general $\sigma$-model action by substituting the equations of motion for $h_{ab}$ into it:
\begin{equation}
 \partial_\mu X_a \partial^\mu X_b = \frac{1}{2} h_{ab} h^{cd} \partial_\mu X_c \partial^\mu X_d
\label{eq:eom_for_worldsheet_metric_h} \nonumber
\end{equation}
We can then invert these equations to find $h_{ab}$ as a function of $\partial_\mu X_a \partial^\mu X_b$, up to an overall rescaling factor (the combination $h_{ab} h^{cd}$ is clearly conformally invariant):
\begin{equation}
 h_{ab} = a(\tau,\sigma) \begin{pmatrix}
                          \dot{X}^2           & \dot{X}^\mu X'_\mu \\
                          \dot{X}^\mu X'_\mu  & X'^2
                         \end{pmatrix}
\label{eq:worldsheet_metric_h_in_a_generic_gauge} \nonumber
\end{equation}
Now, if a worldsheet coordinate transformation $(\tau,\sigma) \to (\tilde{\tau},\tilde{\sigma})$ brings this metric to conformal gauge, i.e. if it makes it diagonal and traceless (the overall scaling factor can then be eliminated by a conformal transformation), then it must satisfy the following set of conditions:
\begin{eqnarray}
 0 & = & h_{00} \left[ \left( \frac{\partial\tau}{\partial\tilde{\tau}} \right)^2 + \left( \frac{\partial\tau}{\partial\tilde{\sigma}} \right)^2
  \right] + 2h_{01} \left[ \left( \frac{\partial\tau}{\partial\tilde{\tau}} \right) \left( \frac{\partial\sigma}{\partial\tilde{\tau}}
   \right) + \left( \frac{\partial\tau}{\partial\tilde{\sigma}} \right) \left( \frac{\partial\sigma}{\partial\tilde{\sigma}} \right) \right] 
    \nonumber \\
   & & + h_{11} \left[ \left( \frac{\partial\sigma}{\partial\tilde{\tau}} \right)^2 + \left( \frac{\partial\sigma}{\partial\tilde{\sigma}}
        \right)^2 \right] \nonumber \\
 0 & = & h_{00} \left( \frac{\partial\tau}{\partial\tilde{\tau}} \right) \left( \frac{\partial\tau}{\partial\tilde{\sigma}} \right) + h_{01}
          \left[ \left( \frac{\partial\tau}{\partial\tilde{\tau}} \right) \left( \frac{\partial\sigma}{\partial\tilde{\sigma}} \right) +
           \left( \frac{\partial\sigma}{\partial\tilde{\tau}} \right) \left( \frac{\partial\tau}{\partial\tilde{\sigma}} \right) \right] 
            \nonumber \\
   &   & + h_{11} \left( \frac{\partial\sigma}{\partial\tilde{\tau}} \right) \left( \frac{\partial\sigma}{\partial\tilde{\sigma}} \right)
\label{eq:conditions_on_coord_transf_NG_to_conf_gauge} \nonumber
\end{eqnarray}
All the required derivatives can be obtained from the first two equations \eqref{eq:equality_of_global_coords_in_two_versions_of_ansatz_K_JJ}, and then it is just a matter of algebra to check that, as expected, these equations are verified.

At this point, it is also interesting to notice that the standard string closedness constraint $X_\mu (\tau,\sigma + \sigma_0) = X_\mu (\tau,\sigma)$, where $\sigma_0$ is some period, is not gauge invariant. In fact, this solution is a clear example of this, since condition \eqref{eq:closedness_constraint_JJ} ensures that the string is closed at constant global time $t$ in both gauges. While we have $t=\tau$ in Kruczenski's gauge and thus standard closedness holds, this is not the case for Jevicki's and Jin's solution, due to the presence of $f(\tilde{\sigma})$. The gauge-invariant object here is $t$ and thus if we impose the standard closedness constraint in static gauge $t=\tau$, we automatically have closedness at constant $t$ in all gauges, but this translates into closedness at constant $\tau$ only in those gauges in which $t=\mathrm{const}$ implies $\tau = \mathrm{const}$.

\section{Computing $q_2$ for the N-folded GKP and the Kruczenski solutions}
\label{sec:app_computing_qtilde_2_for_N-folded_GKP_and_K}

The easiest way of computing $q_2$ in both cases is by using equation \eqref{eq:qktilde_from_spin_chain_patched_K}, which yields all conserved charges for the patched Kruczenski solution. This solution is discussed in detail in section \ref{sec:patching_K}; here we simply recall that it allows arbitrary angular separations $0 < \Delta\theta_j < \pi$ between each pair of consecutive cusps. As we previously observed in section \ref{sec:spectral_curve_patched_K}, all results concerning the spectral curve of this generalised solution reduce to those obtained for the Kruczenski spiky string if we set $\Delta\theta_j = 2 n \pi/K$, $\forall j$, where $n$ is a natural number counting how many times the Kruczenski string winds around the centre of $AdS_3$. Furthermore, we also saw at the end of section \ref{sec:spectral_curve_JJ} that, by setting $n = K/2$, these results in turn reduce to those associated with the N-folded GKP case, where $2 N = K$. Therefore, we can compute the conserved charge $q_2$ by specialising the general expression \eqref{eq:qktilde_from_spin_chain_patched_K} to the desired simpler case.

We start by evaluating it for $k=2$:
\begin{eqnarray}
 q_2 & = & \frac{4 S^2}{K^2} \sum_{1 \leq j_1 < j_2 \leq K} \sin \left( \frac{\theta_{j_2} - \theta_{j_1}}{2} \right)
  \sin \left( \frac{\theta_{j_1} - \theta_{j_2}}{2} \right) \nonumber \\
 & = & - \frac{4 S^2}{K^2} \sum_{1 \leq j_1 < j_2 \leq K} \sin^2 \left( \frac{1}{2} \sum_{l = j_1 + 1}^{j_2} \Delta\theta_l \right)
\label{eq:computing_qtilde_2_for_K_1}
\end{eqnarray}
where we have used $\theta_m\equiv \sum_{j=1}^m \Delta\theta_j$.

We now specialise to the Kruczenski case, by setting $\Delta\theta_j = 2 n \pi/K$, $\forall j$, which implies:
\begin{equation}
 \sum_{l = j_1 + 1}^{j_2} \Delta\theta_l = \frac{2 n \pi}{K} (j_2 - j_1)
\label{eq:simplification_inside_qtilde_2_in_K_case} \nonumber
\end{equation}
By substituting this into \eqref{eq:computing_qtilde_2_for_K_1} and introducing the new index $m = j_2 - j_1$, which replaces $j_2$, we obtain:
\begin{eqnarray}
 q_2 & = & - \frac{4 S^2}{K^2} \sum_{j_1=1}^{K-1} \sum_{m=1}^{K-j_1} \sin^2 \left( \frac{n \pi}{K} m \right)
  = - \frac{2 S^2}{K^2} \sum_{j_1=1}^{K-1} \sum_{m=1}^{K-j_1} \left[ 1 - \cos \left( \frac{2 n \pi}{K} m \right) \right] \nonumber \\
 & = & - \frac{2 S^2}{K^2} \sum_{j_1=1}^{K-1} \left[ K - j_1 - \sum_{m=1}^{K-j_1} \cos \left( m \frac{2 n \pi}{K} \right) \right]
\label{eq:computing_qtilde_2_for_K_2} \nonumber
\end{eqnarray}
We now use the general result for the Dirichlet kernel:
\begin{equation}
 1 + 2 \sum_{k=1}^n \cos (k x) = \frac{\sin \left[ \left( n + \frac{1}{2} \right) x \right]}{\sin \left( \frac{x}{2} \right)}
\label{eq:general_result_sum_of_cosines}
\end{equation}
to calculate the last remaining sum over $m$:
\begin{eqnarray}
 q_2 & = & - \frac{2 S^2}{K^2} \sum_{j_1=1}^{K-1} \left\{ K - j_1 - \frac{1}{2} \left[ \frac{\sin \left[ \left( K - j_1 + \frac{1}{2} \right) 
  \frac{2 n \pi}{K} \right]}{\sin \left( \frac{n \pi}{K} \right)} - 1 \right] \right\} \nonumber \\
 & = & - \frac{2 S^2}{K^2} \sum_{j_1=1}^{K-1} \left\{ K - j_1 + \frac{1}{2} - \frac{1}{2} \frac{\sin \left(  \frac{n \pi}{K}
  - j_1 \frac{2n \pi}{K} \right)}{\sin \left( \frac{n \pi}{K} \right)} \right\} \nonumber \\
 & = & - \frac{2 S^2}{K^2} \sum_{j_1=1}^{K-1} \left\{ K - j_1 + \frac{1}{2} - \frac{1}{2} \left[ \cos \left( j_1 \frac{2 n \pi}{K}
  \right) - \cot \left( \frac{n \pi}{K} \right) \sin \left( j_1 \frac{2 n \pi}{K} \right) \right] \right\} \nonumber \\
 & = & \frac{S^2}{K^2} \left[ - 2 \left( K + \frac{1}{2} \right) (K-1) + (K-1) K + \sum_{j_1=1}^{K-1} \cos \left( j_1 
  \frac{2 n \pi}{K} \right) \right. \nonumber \\
 & & - \left. \cot \left( \frac{n \pi}{K} \right) \sum_{j_1=1}^{K-1} \sin \left( j_1 \frac{2 n \pi}{K} \right) \right]
\label{eq:computing_qtilde_2_for_K_3} \nonumber
\end{eqnarray}
At this point, we can evaluate the sums over $j_1$ by using \eqref{eq:general_result_sum_of_cosines} and the analogous result:
\begin{equation}
 \sum_{k=1}^n \sin (k x) = \frac{\sin x + \sin (nx) - \sin[(n+1) x]}{2(1 - \cos x)}
\label{eq:general_result_sum_of_sines} \nonumber
\end{equation}
thus obtaining:
\begin{eqnarray}
 q_2 & = & - S^2 + \frac{S^2}{K^2} + \frac{S^2}{2 K^2} \left\{ \frac{\sin \left[ \left( K - \frac{1}{2} \right) \frac{2 n \pi}{K}
  \right]}{\sin \left( \frac{n \pi}{K} \right)} - 1 \right\} \nonumber \\
 & & - S^2 \cot \left( \frac{n \pi}{K} \right) \frac{\sin \left( \frac{2 n \pi}{K} \right) + \sin \left[ (K-1) \frac{2 n
  \pi}{K} \right] - \sin (2 n \pi) }{2K^2 \left[ 1 - \cos \left( \frac{2 n \pi}{K} \right) \right]} \nonumber\\
 & = & - S^2
\label{eq:computing_qtilde_2_for_K_4} \nonumber
\end{eqnarray}
Since the result is independent of $n$, it also holds for the N-folded GKP case.

One may wonder whether a similar relation exists for the patched Kruczenski solution, but the answer is negative: as a counter-example, we study the case $K=3$, with $\Delta\theta_1 = 5\pi/6$, $\Delta\theta_2 = 2\pi/3$, $\Delta\theta_3 = \pi/2$ and consequently $n = 1$. It is easy to check from \eqref{eq:computing_qtilde_2_for_K_1} that:
\begin{equation}
 q_2 = - S^2 \frac{4}{9} \left[ \sin^2 \left( \frac{\Delta\theta_2}{2} \right) + \sin^2 \left( \frac{\Delta\theta_2 + \Delta\theta_3}{2}
  \right) + \sin^2 \left( \frac{\Delta\theta_3}{2} \right) \right] = - S^2 \frac{7 + \sqrt{3}}{9}
\label{eq:counter-example_for_patched_K} \nonumber
\end{equation}
The only property that continues to hold for the patched solution is the fact that $q_2 < 0$, as we can easily see from \eqref{eq:computing_qtilde_2_for_K_1}.

\end{document}